\documentclass[twocolumn]{aastex62}
\usepackage{float}
\usepackage{textcomp}
\usepackage{graphicx}
\usepackage{enumitem}
\usepackage{comment}
\usepackage{makecell}
\usepackage{environ}
\usepackage{multirow}
\usepackage{orcidlink}
\usepackage{makecell}
\usepackage{array}
\usepackage{tabularx}
\usepackage{lineno}

\def\sref{\S \ref}

\received{21 Jul., 2023}
\revised{23 Feb., 2024}
\accepted{1 Mar., 2024}
\submitjournal{ApJ}

\usepackage{xcolor} 
\usepackage{soul}   

\shortauthors{Adhikari et al.}

\begin{document}
\title{Constraining the PG 1553+113 binary hypothesis: interpreting hints of a new, 22-year period}

\correspondingauthor{Sagar Adhikari}
\email{E-mails: sagara@clemson.edu, ppenil@clemson.edu}
\email{majello@clemson.edu}

\author{S. Adhikari\orcidlink{0009-0006-1029-1026}}
\affil{Department of Physics and Astronomy, Clemson University,	Kinard Lab of Physics, Clemson, SC 29634-0978, USA}

\author{P. Pe\~nil\orcidlink{0000-0003-3741-9764}}
\affil{Department of Physics and Astronomy, Clemson University,	Kinard Lab of Physics, Clemson, SC 29634-0978, USA}

\author{J.R. Westernacher-Schneider\orcidlink{0000-0002-3047-7200}}
\affil{Leiden Observatory, Leiden University, P.O. Box 9513, 2300 RA Leiden,
The Netherlands}

\author{A. Dom\'inguez\orcidlink{0000-0002-3433-4610}}
\affil{IPARCOS and Department of EMFTEL, Universidad Complutense de Madrid, E-28040 Madrid, Spain}

\author{M. Ajello\orcidlink{0000-0002-6584-1703}}
\affil{Department of Physics and Astronomy, Clemson University,	Kinard Lab of Physics, Clemson, SC 29634-0978, USA}

\author{S. Buson\orcidlink{0000-0002-3308-324X}}
\affil{Julius-Maximilians-Universit\"at, 97070, W\"urzburg, Germany}

\author{A. Rico\orcidlink{0000-0001-5233-7180}}
\affil{IPARCOS and Department of EMFTEL, Universidad Complutense de Madrid, E-28040 Madrid, Spain}

\author{J. Zrake\orcidlink{0000-0002-1895-6516}}
\affil{Department of Physics and Astronomy, Clemson University,	Kinard Lab of Physics, Clemson, SC 29634-0978, USA}

\begin{abstract}
PG 1553+113 is a well-known blazar exhibiting evidence of a $\sim\! 2.2$-yr quasi-periodic oscillation (QPO) in radio, optical, X-ray, and $\gamma$-ray bands.
Since QPO mechanisms often predict multiple QPOs, we search for a second QPO in its historical optical light curve covering a century of observations.
Despite challenging data quality issues, we find hints of a $21.8 \pm 4.7$ yr oscillation.
On its own, this $\sim\! 22$-yr period has a modest statistical significance of $1.6\sigma$ when accounting for the look-elsewhere effect.
However, the joint significance of both the $2.2$- and $22$-yr periods arising from colored noise alone is $\sim 3.6\sigma$.
The next peak of the 22-yr oscillation is predicted to occur around July 2025. We find that such a $\sim\,$10:1 relation between two periods can arise in the gas dynamics of a plausible supermassive black hole binary model of PG 1553+113. While the 22-yr QPO is preliminary, an interpretation of PG 1553+113's two QPOs in this binary model suggests that the binary engine has a mass ratio $\gtrsim 0.2$, an eccentricity $\lesssim 0.1$, and accretes from a disk with characteristic aspect ratio $\sim 0.03$. The putative binary radiates nHz gravitational waves, but the amplitude is $\sim10-100$ times too low for detection by foreseeable pulsar timing arrays.
\end{abstract}

\section{Introduction} \label{sec:intro}
Blazars are active galactic nuclei whose relativistic and collimated jets are closely aligned with our line of sight \citep[e.g.,][]{wiita_lecture}. The non-thermal radiation from the jet is relativistically enhanced to the point where it easily outshines the entire host galaxy.

In addition to their spectacular energy output, blazars commonly show flux variability across the entire electromagnetic spectrum \citep{urry1996overview} over a vast range of time scales \citep[e.g.,][]{jurkevich1971light, miller1989detection, wagner1995intraday, lainela199965, kranich1999evidence}.

Blazar variability is commonly associated with stochastic processes
\citep[e.g.,][]{tavecchio2020distribution}. Yet, some blazars exhibit evidence of quasi-periodic oscillations \citep[QPOs, e.g.,][]{wiita2011quasi, prokhorov_set}. Among them, PG 1553+113 is one of the most studied \citep[e.g.,][]{ackermann2015multiwavelength, caproni2017jet, agarwal2021multi}. This system is found to exhibit periodic flux variations in $\gamma$-ray, X-ray, optical, and radio frequencies, with a $\sim$2.2-yr period and with high local significance
\citep[e.g. $2.8\sigma$ trial-corrected significance in the optical band,][]{penil_mwl_pg1553}.
This has motivated several authors to interpret PG 1553+113's $2.2$-yr period as due to a binary system of supermassive black holes (SMBHs); for example, \citet{caproni2017jet} and \citet{huang+2021} discuss jet precession scenarios involving binaries.

Moreover, a recent study by \citet{penil_mwl_pg1553} reveals that the light curves (LCs) of PG 1553+113 exhibit long-term ($>$5 yr) trends of flux increase or decrease, spanning $\sim$15 yrs and covering radio, optical, X-ray, and $\gamma$-ray bands. These long-term trends are observed alongside 2.2-yr oscillations in the LCs. Specifically for the optical band, a decrease in the optical flux from 2009 to 2015 and an increase from 2015 to 2021 is reported. If these trends are segments of a longer-term period, it could help to distinguish between single and binary models of PG 1553+113.

Indeed, long-term variability in blazar emission can be explained within single or binary SMBH scenarios. For example, an analysis of a blazar's QPOs by \citet{sarkar2021multiwaveband} concludes that the most likely explanation for an observed $\sim$year-long decay of their reported QPO is a curved jet. On the other hand, \citet{dey2018authenticating} report a double periodicity in OJ 287, with a 5:1 ratio between the two periods. In that system, the shorter period is thought to be associated with the orbital period of an extreme mass ratio binary black hole that is inclined with respect to the accretion disk, while the longer period is associated with relativistic precession of the binary orbit. On the other hand, for binaries with nearly equal masses and no inclination with respect to the disk, a second period might still occur due to a different mechanism, a so-called ``lump," referring to an orbiting overdensity in the circumbinary disk that modulates the accretion rate to the black holes on a time scale of $5-10$ binary orbits \citep[e.g.][]{macfadyen+2008, shi+2012, noble+2012, dorazio+2013, farris+2014, bowen_lump, munoz+2020, duffell+2020, zrake+2021}. Detection of this longer-term modulation, in addition to the shorter one associated with the binary orbital period, could suggest constraints on binary SMBH scenarios.

In this work, we study the variability of PG 1553+113 on multi-decade time scales using optical data going back roughly 100 years.
This paper is organized as follows. We discuss the data access in $\S$\ref{sec:data_access}, explain the methodology used for the periodicity analysis in $\S$\ref{sec:methodology}, provide an overview of the impact on our periodicity analysis from gaps in the LC in $\S$\ref{sec:gap_impact}, present our results in $\S$\ref{sec:results}, and discuss the constraints the binary SMBH hypothesis in $\S$\ref{sec:discussion}. We conclude in $\S$\ref{sec:summary}.
\section{Data Sample} \label{sec:data_access}
To constrain potential longer-term ($>$10-yr) periodic emission, we need time-series data spanning about a century. To accomplish that, we make use of several publicly accessible databases.
\subsection{DASCH}
The historical optical observations in our study are taken from the Digital Access to a Sky Century @ Harvard (DASCH) database, which provides an excellent archive of optical observations from the 20th century \citep[][]{dasch_database}. The DASCH  project includes scanned photographic plates from a network of optical telescopes (e.g., A Series, ADH Series) belonging to Harvard University, covering both celestial hemispheres from 1885 to 1992.\footnote{\url{http://dasch.rc.fas.harvard.edu/telescopes.php}} DASCH provides data in the Johnson B and V bands for all sources calibrated using SExtractor and \emph{astrometry.net}, ensuring photometric consistency between observations from multiple plates \citep[][]{dasch_database}.
Since the magnitude data is digitized from photographic plates, there is a number of quality issues with the scanned data points.\footnote{\url{http://dasch.rc.fas.harvard.edu/database.php##AFLAGS_ext}} For this study, we use data with uncertainties less than $\pm 0.4$ magnitude, which is a default filter of the DASCH database.
We employ data from the V-band since more observatories provide data in this band. DASCH provides 312 data points from 1900 until 1992 for PG 1553+113.
\subsection{Complementary databases}\label{sec:databases}
To have a more complete and extended coverage of the optical emission from PG 1553+113, we also use data from more recent optical surveys, which are CSS \citep[Catalina Sky Survey, ][]{drake2009}\footnote{\url{http://nesssi.cacr.caltech.edu/DataRelease/}}, ASAS-SN \citep[All-Sky Automated Survey for Supernovae, ][]{shappee2014,kochanek2017}\footnote{\url{http://www.astronomy.ohio-state.edu/asassn/index.shtml}}, AAVSO (American Association of Variable Star Observers)\footnote{\url{http://https://www.aavso.org/data-download/}} International Database, and ZTF \citep[Zwicky Transient Facility, ][]{zwicky_observatory}.\footnote{\url{https://www.ztf.caltech.edu/}} These databases provide 353, 225, 90, and 401 data points, respectively. The combination of V-band data from these observatories provides optical observations from 2005 to 2022 for PG 1553+113. The combined data from these contemporary databases are hereafter referred to as Modern Databases (MDB).
\subsection{Photometric calibration}
To ensure compatibility between the DASCH and the MDB LCs, we compare a constant magnitude source across the databases. As a single object observed in all databases is not found, the magnitude of a few sources observed in multiple databases is compared. Specifically, V* W Vir and V* FI Psc each are found to have observations in three of the five databases used for PG 1553+113.

V* W Vir is a well-studied cepheid variable \citep[e.g.,][]{sandage2006absolute, madore2013preliminary}, and we find that the median magnitude of the DASCH LC is consistent with the AAVSO and ASAS-SN counterparts within the standard deviation. The medians and standard deviations are 9.98, 9.90, 9.92, and 0.60, 0.51, 0.45, respectively, for the three databases.
For V* FI Psc, which is an RR Lyrae variable observed in CSS, ZTF, and ASAS-SN \citep[e.g.,][]{drake2013probing}, we find that the median magnitudes are also within the standard deviations with medians 13.48, 13.46, 13.63, and standard deviations 0.29, 0.28, 0.34, respectively.

Since the photometry of DASCH, AAVSO, and ASAS-SN are compatible without any offset correction, similar to CSS, ZTF, and ASAS-SN, we conclude that the databases are in photometric calibration with each other, and the LCs can be combined without correction.
\section{Methodology} \label{sec:methodology}
\subsection {Pre-analysis procedure}
Combining all the LCs from DASCH and MDB, we get time coverage from 1900 to 2022. However, prior to our data analysis, we apply several pre-processing steps to the data. Optical data in the DASCH database were digitized from plates that were stored for extended periods and originated from outdated data acquisition technologies. These factors result in large uncertainties in the recorded data. Thus, we exclude data with a signal-to-noise ratio lower than three \citep[e.g.,][]{otero2023}, and we filter any data taken before 1920 as they showed a large dispersion from the mean magnitude of 14.02 mag that is unlikely to be physical. These quality constraints result in a reduced time coverage of the LC from 1920 to 2022.

The data are grouped into 28-day bins. Binning is essential for us to look beyond the intra-day or even week-long variabilities of the object of interest. The new data points result from the median value of the data points within the binning window. 
Each bin's uncertainty is taken to be the interquartile range of the uncertainties of the data points in that bin, which proves to be suitable for handling noisy datasets \citep[e.g.,][]{bindu_binning, negi_binning}.
This binning provides an adequate compromise between a computationally manageable analysis and a sensitivity to long-term variations on the order of a year.

\begin{figure*}[htp]
    \centering
    \includegraphics[width=0.8\linewidth]{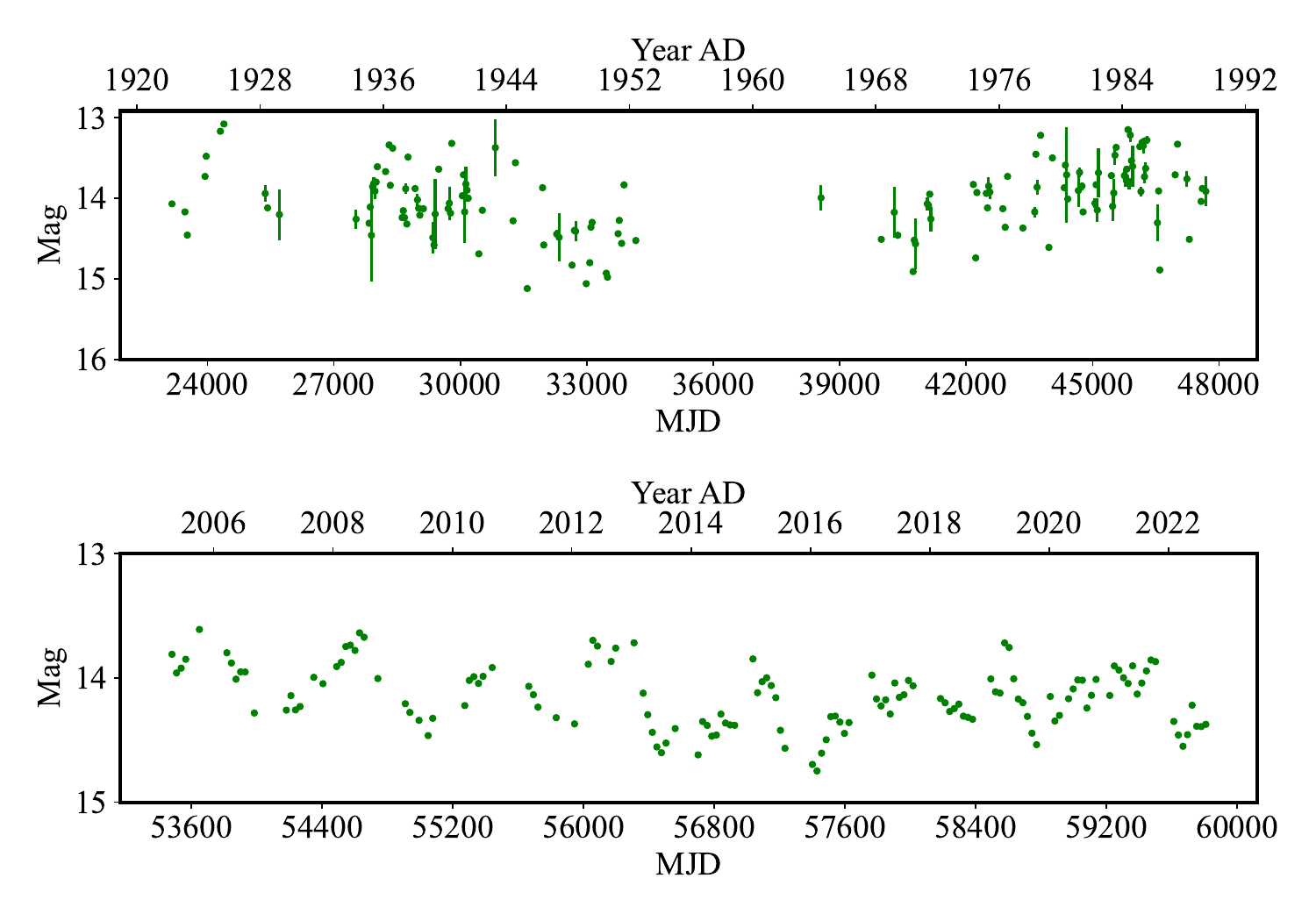}
    \caption{The V-band LCs of PG 1553+113, after applying a 28-day binning to the DASCH data (upper panel) and the modern database data (lower panel).}
    \label{fig:pg53lc}
\end{figure*}

After applying these pre-analysis processes, we notice a gap of $17\%$ on the PG 1553+113 LC. The resulting LC
is shown in Figure~\ref{fig:pg53lc}. The figure shows 286 data points from the DASCH database and MDB, displayed in the upper and lower panels, respectively. We search for periods in the 8- to 40-yr range. The lower bound is selected according to binary lump hypothesis (discussed in $\S$\ref{sec:bhypo}), and the upper bound is chosen based on the limited detectability of periods above roughly half the range of the LC ($\approx$100 yr).
\subsection{Periodicity-search methods}
The unevenly-spaced LC in our study necessitates the use of periodicity-search methods that can robustly manage non-uniform sampling. Thus, we use the Generalised Lomb-Scargle periodogram \citep[GLSP,][]{lomb_1976, scargle_1982, lomb_gen} and the Weighted Wavelet Z-transform \citep[WWZ,][]{foster_wwz}. The GLSP is chosen for its ability to account for the uncertainties in the observation and its suitability for unevenly-spaced LCs \citep[as demonstrated by][]{penil_2020}. On the other hand, the WWZ decomposes a time series in the frequency and time domains. It finds transient and stationary periodic patterns at different timescales. In both methods, we estimate the uncertainty of the observed period using the FWHM of the peak period \citep[e.g.,][]{otero2020quasi}.
\subsection{Generalized Lomb-Scargle periodogram}
The GLSP is a method to analyze periodicity in unevenly spaced time series, combining elements of the Fourier transform and the least-squares statistical method. The Fourier transform provides information on the relative amplitudes of different frequencies present in the data, while the least-squares approach yields the statistical significance of each peak and is adaptable to unevenly spaced data \citep[][]{VanderPlas_2018}.

To calculate the GLSP, a sinusoidal function is used, and the chi-squared values of the observed data are determined, as outlined by \citet{VanderPlas_2018}.
The general sinusoidal fit function is defined as,
\begin{equation}
    y(t;f) = A_f \sin(2\pi f(t-\phi_f)).
\end{equation}
Here, $A_f$ is the amplitude, $f$ frequency, $t$ time, and $\phi_f$ is the phase of the sine wave. A chi-squared statistic is constructed at each frequency: $\chi^2(f) \equiv \sum_n \left(y_n - y(t_n;f)\right)^2$. Here, $n$ in the subscript represents the frequency values for the fit. A best fit model is determined by obtaining the minimum of this chi-squared value, $\hat{\chi}^2(f)$. To be able to handle the uncertainties in each data point, a modified $\chi^2$ value is computed as,
\begin{equation}
    \chi^2(f) \equiv \sum_n \left(\frac{y_n - y(t_n;f)}{\sigma_n}\right)^2.
\end{equation}
Here, $\sigma_n$ represents the uncertainties. The GLSP periodogram as a function of frequency is defined as,
\begin{equation}
    P(f) = \frac{1}{2}[\hat{\chi}_0^2 - \hat{\chi}^2(f)].
    \label{eqn:pgram}
\end{equation}
Here, $\hat{\chi}_0^2$ corresponds to the sinusoidal reference model. This periodogram is normalized and presented in this work using the implementation of the \citet[][]{astropy:2022}.
\subsection{Weighted wavelet Z-transform}
The weighted wavelet Z-transform (WWZ) is a method to analyze the frequencies present in time series data using a sliding Morlet wavelet. A Morlet wavelet is a waveform characterized by a harmonic oscillation with a Gaussian decay profile, usually defined as
\begin{equation}
    f(z) = f(\omega(t-\tau)) = e^{-C\omega^2 (t-\tau)^2}(e^{i\omega(t-\tau)}-e^{-1/4C}),
    \label{eq:morlet}
\end{equation}
where $C$ controls the decay of the wavelet, $\omega$ is the scale factor, $t$ is time, and $\tau$ is the time shift \citep{foster_wwz}. WWZ involves comparing the wavelet model with the time series by varying the frequency of the harmonic component and calculating the weighted variations between the data and the model. The WWZ transform equation is given by:
\begin{equation}\label{eqn:wwz}
    Z = \frac{(N_{\rm eff} - 3)V_y}{2(V_x - V_y)}.
\end{equation}
Here, $V_x$ is the weighted variation of the data, $V_y$ is the weighted variation of the model function, and $N_{\rm eff} > 3$ is the effective number of data projected onto the wavelet window \citep{foster_wwz}.

It is important to note that the $Z$-value is unbounded. We normalize the $Z$-value and represent it by a color bar in WWZ plots. The WWZ implementation of RedVoxInc\footnote{\url{https://github.com/RedVoxInc/libwwz}} is used in this work.

Finally, it is important to consider the impact of edge effects in the wavelet analysis, which is indicated by the cone of influence in wavelet representations. This is a region in time-frequency space where edge effects become significant; the presence of a particular frequency becomes less discernible due to the decrease in the number of data points in the Morlet wavelet. In the case of a finite time series like ours, detecting a particular frequency depends on the number of data points in the sampled frequency curve. As the wavelet approaches the edge, this number decreases, affecting the reliability of the detected frequency or period near the border. We use a white-shaded region indicate the cone of influence in the WWZ plots in this work.
\subsection{Significance levels}
Periodicity searches are limited in large part by noise. Many astrophysical sources (galactic and extragalactic) show erratic brightness fluctuations with steep power spectra, known as red noise 
\citep{papadakis1993improved, vaughan2003characterizing, zhu2016using}. In this context, noise is defined as random variations in the source emission. For a periodicity analysis to be effective, assessing the significance of the frequencies present in time-series data is necessary, which can be achieved through the simulation of artificial LCs. Various techniques have been proposed to model LCs, such as SARIMA and CARMA \citep[see, e.g.][]{feigelson2018autoregressive,tarnopolski2020comprehensive}. SARIMA is best suited for regular time series. Regarding CARMA, it can be applied to uneven time series data. However, the application of this technique to PG 1553+113's historical optical LC presents problems. Specifically, our data exhibit a highly unevenly-distributed nature, characterized by irregular sampling intervals and significant dispersion. CARMA models rely on the estimation of parameters based on the time intervals between observations. When the time series is highly uneven, with large gaps between data points, it becomes difficult to accurately estimate these parameters. To manage a significant dispersion, the CARMA model might need to be more complex. However, a more complex model can lead to overfitting, where the model captures the noise in the data rather than the underlying process.

To address these challenges, we follow an alternative and wide-use method whereby we simulate 100,000 artificial LCs to properly derive the significance of the detected periods and estimate the likelihood of false positives. The artificial LCs are generated following the technique of \citet{emma_lc} using the Python implementation of \citet[][]{connolly_code}, resulting in LCs with the same power spectral density (PSD), probability density function, and sampling as real blazar LCs. In our analysis, to generate artificial LCs, we consider two PSD models: power law \citep[PL, e.g.,][]{ackermann2015multiwavelength} and bending power law (BPL) models.
The resulting periodograms are analyzed to determine the confidence levels of their peaks, calculated based on the percentiles of the power for each period bin in the periodograms.
\subsection{Power spectral density estimation}\label{sec: psd-estimation}
Traditionally, the noise is classified according to the PL index $\beta$ of the PSD \citep[$A*\nu^{-\beta}$ where $\nu$ is the frequency ($\mathrm{yr}^{-1}$), and $A$ is the normalization,][]{rieger_2019}. Other authors suggest using a BPL to fit the PSD, since this approach provides a more realistic model of blazar variability on all time scales \citep{chakraborty_bending_power_law}. Thus, we estimate the significance of the periodicity search using both PSD models. In the case of the BPL, we employ the expression:
\begin{equation} \label{eqn:bending} 
  P(\nu) = A \left( 1 + \left\{ \frac{\nu}{\nu_{b}} \right\}^{\beta} \right)^{-1}, 
\end{equation} 
where $A$ is the normalization, $\beta$ is the spectral index,
and $\nu_{b}$ is the bending frequency \citep{chakraborty_bending_power_law}. 

The parameters of each PSD model ($A$ and $\beta$ for PL and $A$, $\beta$, and $\nu_{b}$ for BPL) are estimated using maximum likelihood and Markov chain Monte Carlo simulations \citep[ML-MCMC,][]{2013PASP..125..306F}:
\begin{itemize}
    \item PL: $A$=0.02$\pm$0.01, $\beta$=0.43$\pm$0.04
    \item BPL: $A$=0.09$\pm$0.01, $\beta$=1.13$\pm$0.09, $\nu_{b}$=0.12$\pm$0.02
\end{itemize}

\section{Impact of Gaps in the Periodicity Analysis}\label{sec:gap_impact}
In the archival optical LCs, as we have for PG 1553+113, we usually find irregular gaps due to a lack of observation during some intervals. Binning the LC smooths out the smaller gaps, but larger gaps remain, and could affect both the peak period in the signal and its significance. We perform a study to understand the impact of gaps in the LC. We want to know how introducing gaps in a synthetic periodic LC affects the value of the detected period, its significance, and whether gaps in a random signal with no period can cause false period detection. For both periodic and noise-dominated cases, the LCs we simulate span 100 yrs, and the periodicity analysis is done with GLSP in the 8- to 40-yr range.

For each case described below, we simulate $N_{\rm sim} = 10^5$ LCs with artificial measurement uncertainties, randomly selected from a normal distribution, for all data points on each LC.
\subsection{Gaps in a periodic light curve}\label{sec:gap_periodic}
\begin{figure}[htp]
    \centering    \includegraphics[width=\columnwidth]{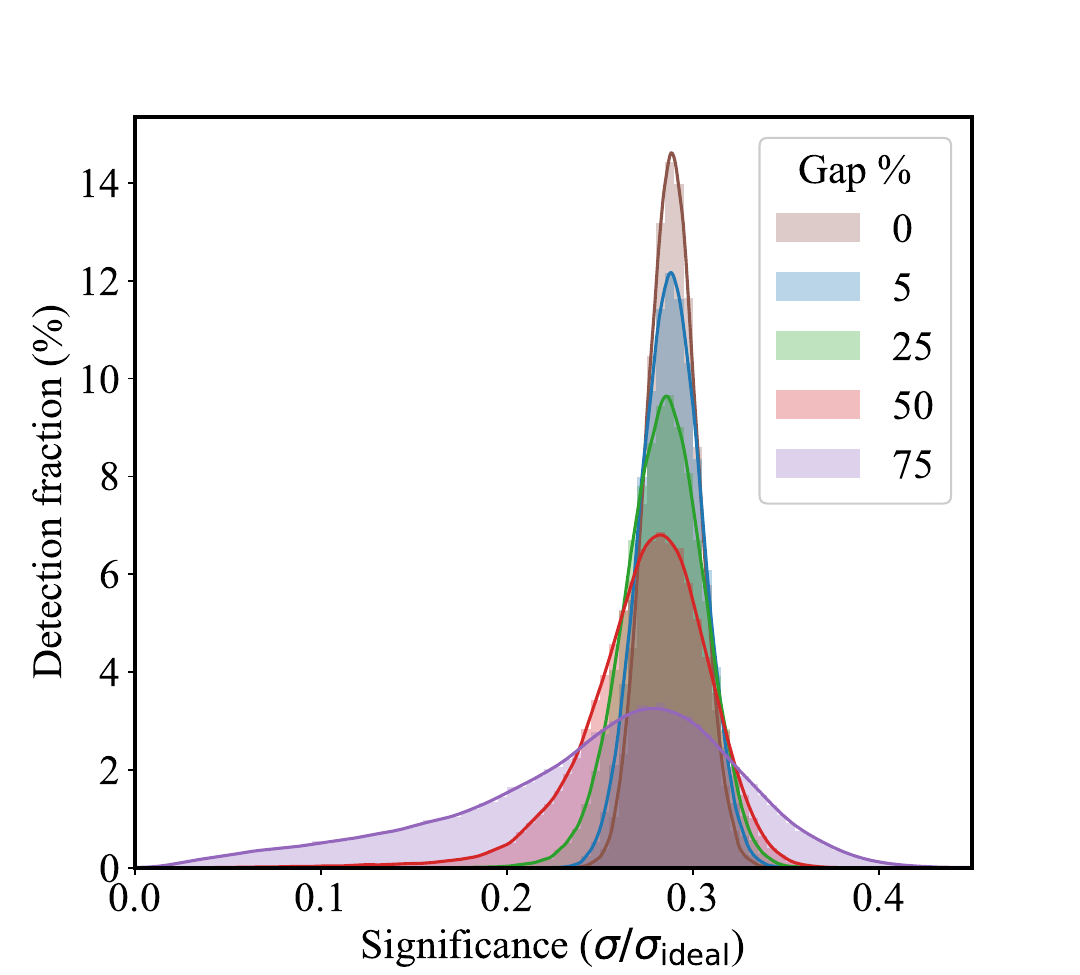}
    \caption{Distribution of the fraction of simulated LCs observed with a relative significance of $\sigma / \sigma_\mathrm{ideal}$ for different gap rates on periodic LCs. The distribution flattens and broadens significantly as the gap fraction increases.
    }
    \label{fig:sine_hist_sigma}
\end{figure}

We simulate $N_\mathrm{sim}$ periodic LCs having four and a half cycles (4.5 cycles of a 22-yr period spanning a 100-yr LC, motivated by the result of this work, see $\S$\ref{sec:results}). For each simulated signal, the period's phase is chosen from a uniform distribution between $-\pi/2$ and $+\pi/2$, since there is no \emph{a priori} expectation for the phase. The intrinsic variability is modeled by a normal distribution $N(0,\sigma)$, where the standard deviation is obtained from the original LC.
A noise-free periodic signal with 4.5 cycles and no uncertainty is taken as a reference to compare with the significances obtained from simulated LCs. The significance of the highest peak in the reference LC's periodogram is $\sigma_\mathrm{ideal}$.
We remove chunks of data randomly from the simulated LCs to resemble the gaps in the PG 1553+113 LC from Figure~\ref{fig:pg53lc}. For instance, ``0\% gap'' means the simulated LCs have noise and uncertainties for each data point but no gaps and for ``25\% gap", we take chunks of 10\% out twice and take 5\%, to resemble a LC with large contiguous gaps.
%
In Figure~\ref{fig:sine_hist_sigma}, the vertical axis shows the detection fraction ($\mathrm{N_{detect} / N_{sim}} \times 100$, where $\mathrm{N_{detect}}$ is the number of periodograms with a specific value of relative significance for the highest peak frequency). The figure shows the fraction of simulated LCs with a given value of relative significance going down as gaps are introduced, compared to the no gaps case. In the case of a genuine periodic signal, the gaps do not significantly change the median of the relative significance (e.g., change in the median of 10\% occurs for 75\% of gaps). However, the distribution of significance broadens, developing a tail towards a lower value. The standard deviation of the significance results changes dramatically (153\% for 50\% of gaps and 440\% for 75\% of gaps). It is evident that for a periodic signal, gaps $<$ 50\% have minimal impact on peak significance.

The fraction of periodic signals with a given significance reduces dramatically with the increase in gaps. For instance, a signal with a 25\% gap is $\sim$50\% less likely to be observed with a given relative significance than the signal without gaps (see Figure~\ref{fig:sine_hist_sigma}).
%
%
%
\subsection{Gaps in a pure-noise light curve}
We simulate white noise and red noise light curves from PG 1553+113-like PSD models to study the impact of gaps.
\subsubsection{Gaps in a white-noise light curve}\label{sec:whitenoise}
\begin{figure}
    \centering    \includegraphics[width=\columnwidth]{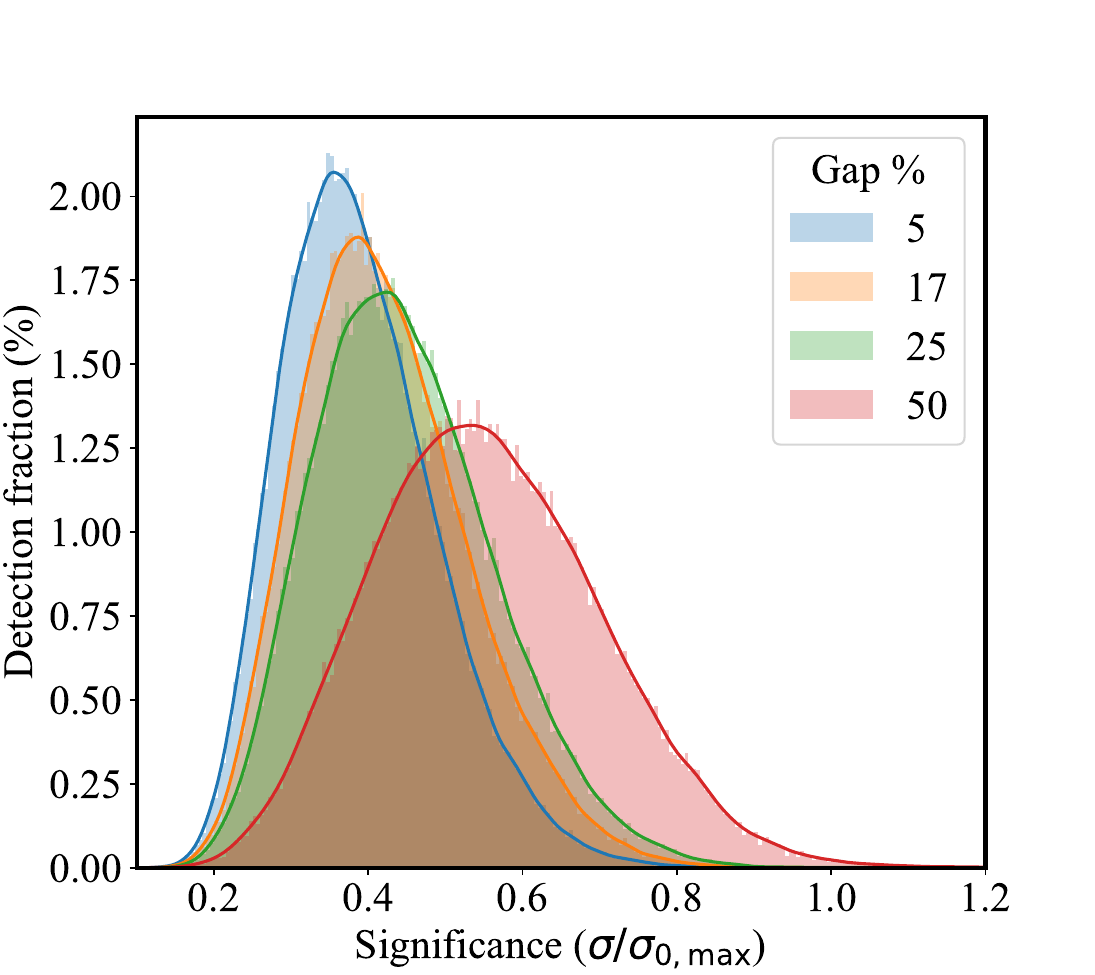}
    \caption{Distribution of the fraction of simulated LCs observed with a relative significance of $\sigma / \sigma_\mathrm{0, max}$ for different gap rates on white-noise LCs. The distribution flattens, broadens, and shifts towards higher significance as the gap fraction increases.
    }
    \label{fig:random_sigma_hist}
\end{figure}
We simulate $N_\mathrm{sim}$ white noise LCs 
as described in $\S$\ref{sec:gap_impact}. We also introduce gaps, just like in the case of the periodic LCs. Figure~\ref{fig:random_sigma_hist} shows the distribution of the relative significance of periods inferred from the random-noise LCs with different amounts of gaps.
Here, $\sigma_{\rm 0, max}$ is defined as the maximum significance across all simulated LCs of any observed period before introducing any gap. We use $\sigma_{\rm 0, max}$, a constant value, to normalize the significance distribution seen in the figure.

The median of relative significance increases with increased gaps.
A gap of $17\%$ increases the median of the relative significance by $\sim$12\% compared to the LC without gaps. For 50\% of gaps, the median of the relative significance is $\sim$50\% larger. It is evident from the figure that gaps can increase the significance and produce false periodicities with higher significances; however, the number, or the likelihood, of such false detections goes down with increased gaps.
\subsubsection{Gaps in a PG 1553+113-like red noise light curve}
\begin{figure}[htp]
    \centering
    
        \includegraphics[width=0.35\textwidth]{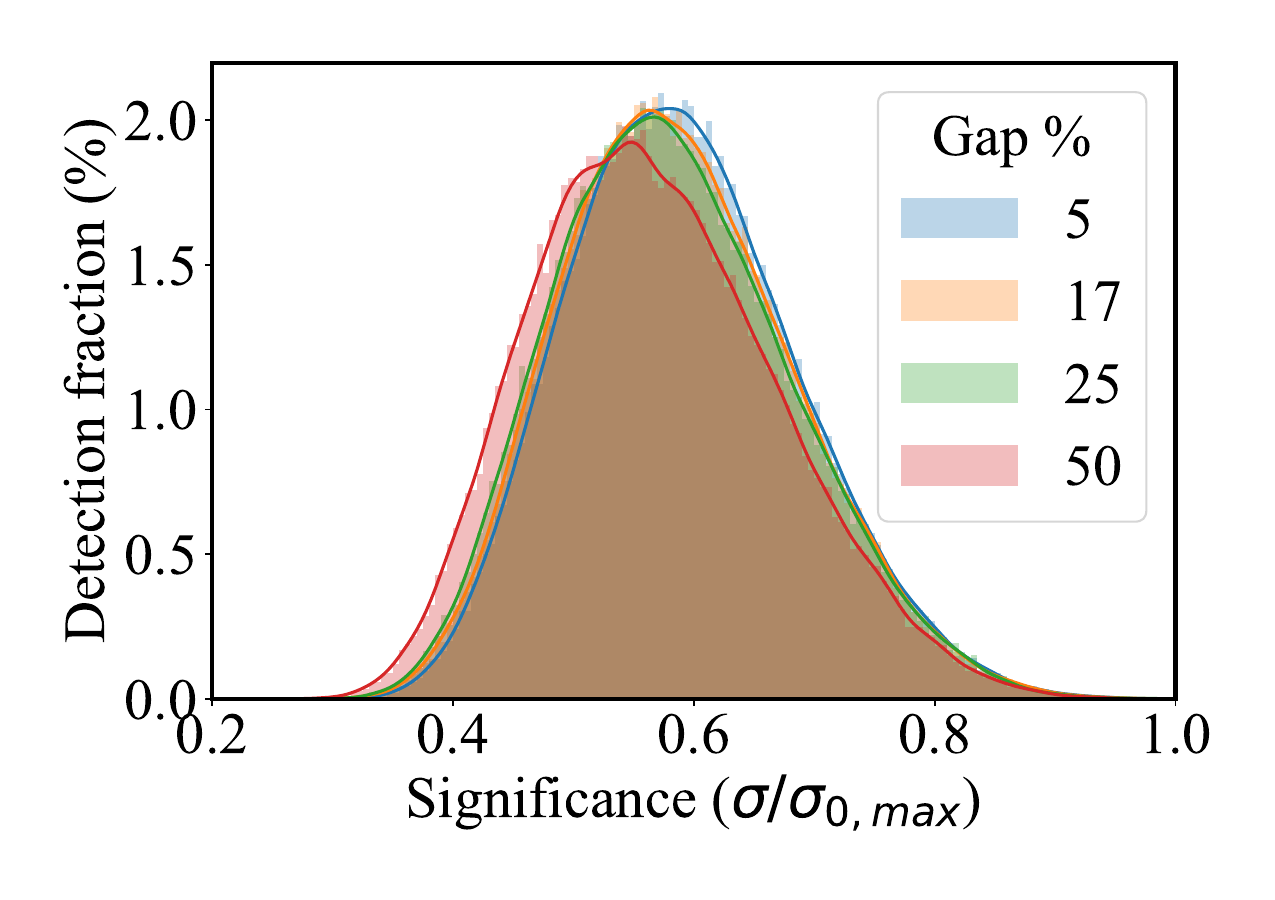}
        
    \hfill
    
        \includegraphics[width=0.35\textwidth]{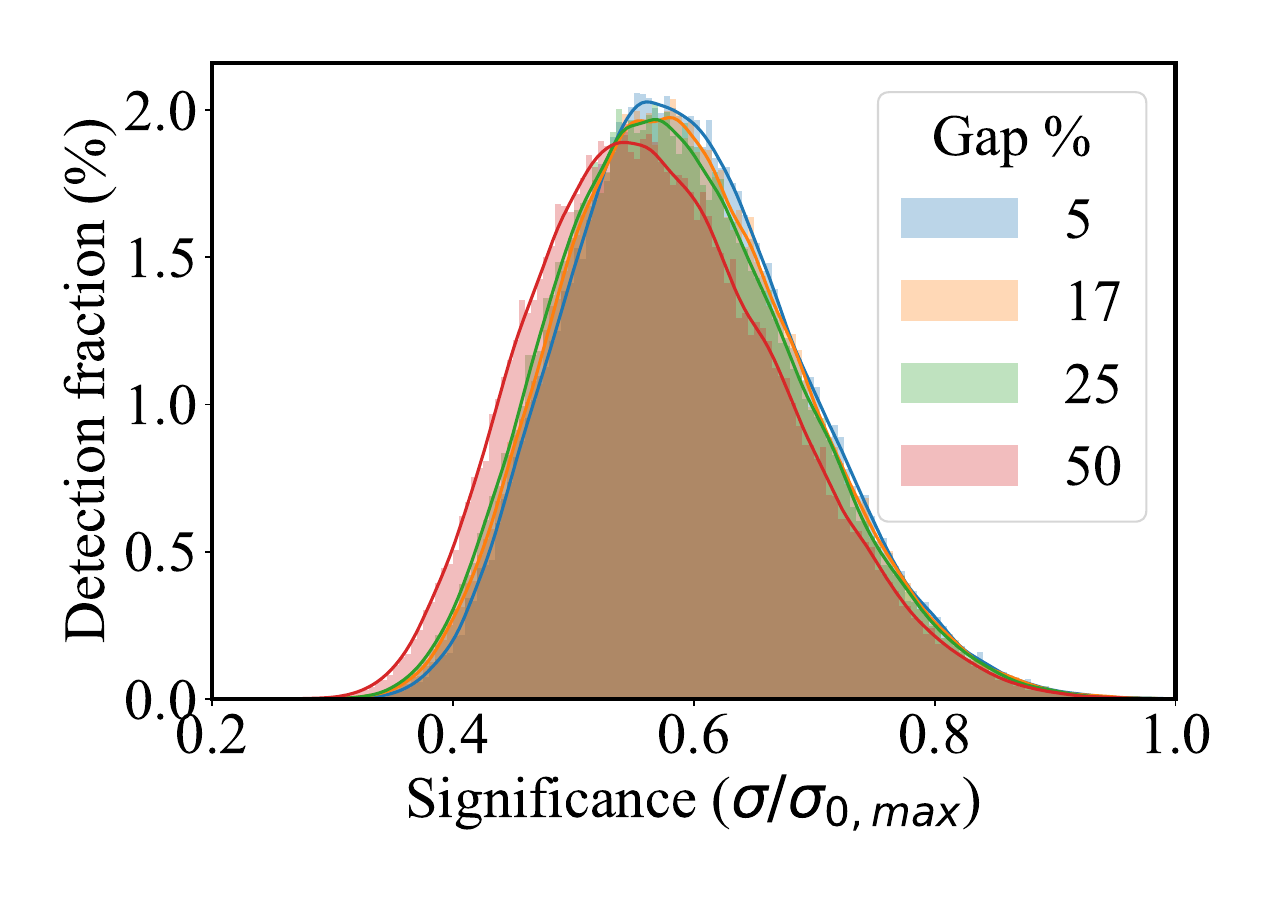}
        
    \caption{Distribution of the fraction of simulated LCs observed with a relative significance of $\sigma / \sigma_\mathrm{0, max}$ for different gap rates in the PL (top panel) and BPL (bottom panel) random-noise LCs. The distribution shifts slightly to the \textbf{left} as the gap fraction increases.
    }
    \label{fig:random_sigma_hist_pl_bpl}
\end{figure}

We simulate $N_\mathrm{sim}$ red-noise LCs along with random uncertainties for each data point on each LC and gaps as described in $\S$\ref{sec:whitenoise}. The red noise is simulated using the methods of \citet{emma_lc} with the model parameters from $\S$\ref{sec: psd-estimation}.
We find that the gaps in the LCs decrease the significance reported for the peak period. We find the median decrease is 1.1\% for the $17\%$ gap case and is 4.5\% for the 50\% gap using the PL model; for the BPL model, we also find a decrease of 1.1\% and a decrease of 4.5\%, respectively (see Figure \ref{fig:random_sigma_hist_pl_bpl}). Additionally, the shift of the histogram to the left as we introduce gaps in the case of PG 1553+113-like LCs closely resembles the periodic case in Figure \ref{fig:sine_hist_sigma}, rather than the random case in Figure \ref{fig:random_sigma_hist}.

\begin{figure}[ht]
    \centering    \includegraphics[width=\columnwidth]{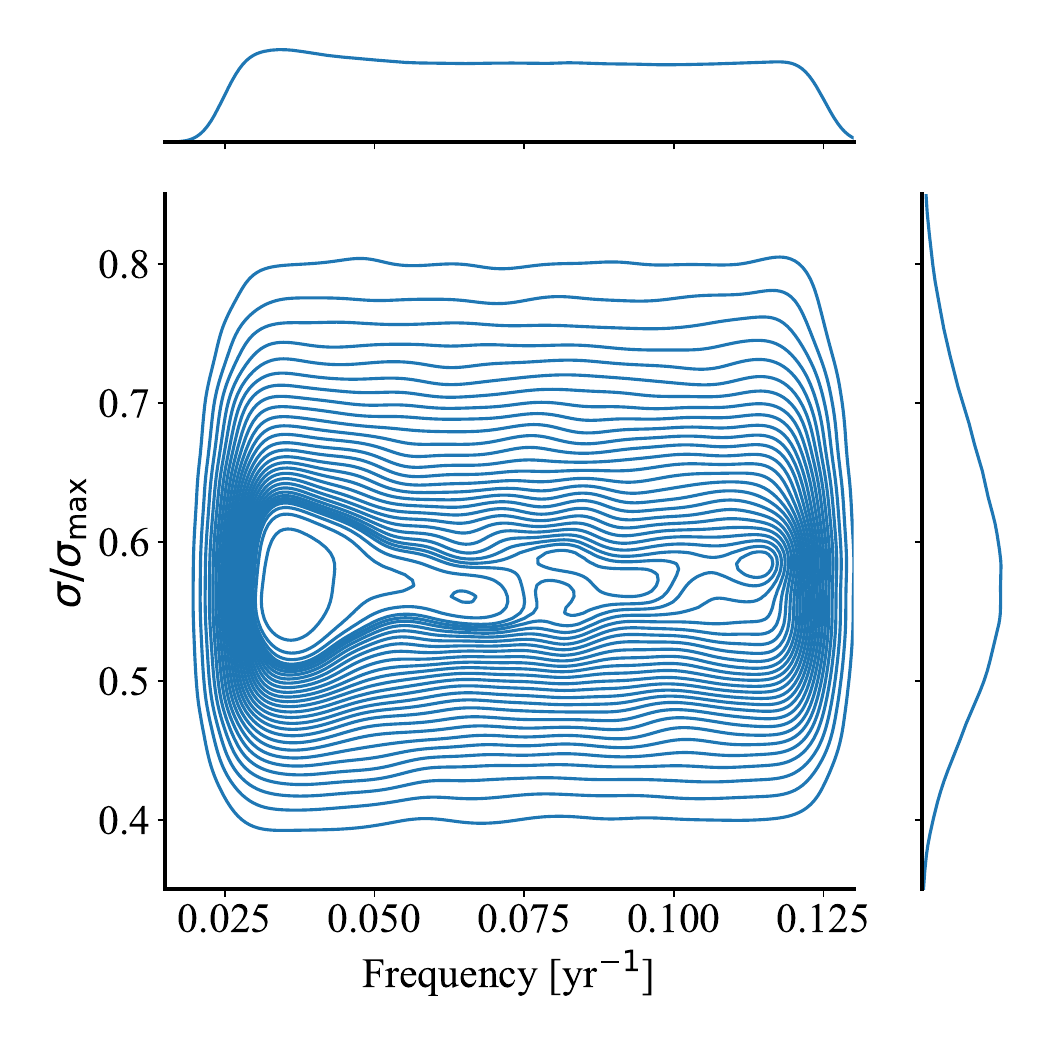}
    \caption{Density plot of peak frequencies with their corresponding significances (relative to the maximum significance) for a PG 1553+113-like BPL random-noise signal with 17\% gap. No preference for any particular period is seen from the main plot as well as the histogram on top.}
    \label{fig:2d_hist_17p}
\end{figure}

Figure~\ref{fig:2d_hist_17p} shows the observed frequency density and the corresponding relative significance of the $N_\mathrm{sim}$ simulated random LCs, with a 17\% gap. The vertical axis shows the significance of a peak frequency relative to the maximum significance and the horizontal axis shows the frequency in the unit of per year. It is clear that a random noise signal does not generate any preferred period (see the constant distribution on top of Figure~\ref{fig:2d_hist_17p}, and ignore the obvious edge effect). The same results are obtained for any amount of gaps. The difference in relative significance values is $\sim 3\%$, and the median frequencies varied by $\sim 2\%$ across all frequency bins. Therefore, the gaps are unlikely to produce a biased generation of a specific period.

We find similar results (relative significances and median frequencies vary by 5\% and 1\%) for PL and (by 3.5\% and 1\%) BPL using the specific PSD models for PG 1553+113 explained in $\S$\ref{sec: psd-estimation}.

In conclusion, the gaps in random noise LCs affect the periodicity analysis, with a potential generation of significant false periodicity with gap rates far exceeding the 17\% rate of PG 1553+113 (e.g.~$\geq\,$50\% of gaps). But this effect is not present when a periodic signal is real. Most importantly, the simulated LCs with the PG 1553+113-like PSD models do not show any effect of gaps in the significance. According to our tests, 17\% of gaps in the optical LC of PG 1553+113 do not introduce any significant evidence that the observed period results from an artifact, nor does it affect the significance of said period.
\section{Results} \label{sec:results}
\begin{figure}[htp]
    \centering    \includegraphics[width=\columnwidth]{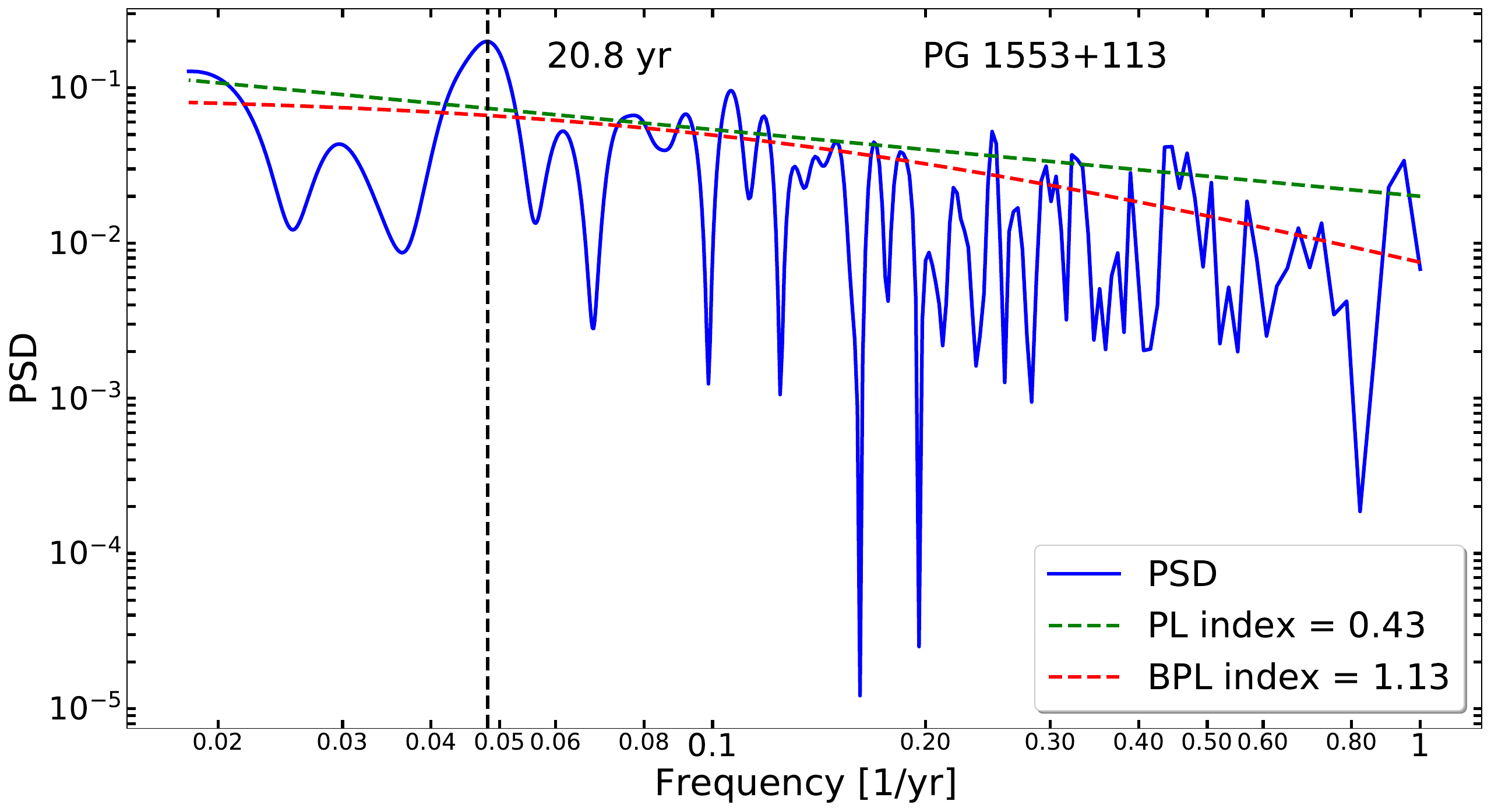}
    \caption{PSD fits according to the PL and BPL estimates. The vertical dotted line shows the position of the peak period at 20.8 yr.}
    \label{fig:psd_estimations}
\end{figure}

The PSD of the PG 1553+113 LC, and its PL and BPL fits from $\S$\ref{sec: psd-estimation}, are shown in Figure~\ref{fig:psd_estimations}. We employ the concept of relative likelihood of models (RLM) to compare these two PSD models. This comparison is conducted by  calculating $\exp\left( \frac{1}{2}(\text{AIC}_{\text{BPL}} - \text{AIC}_{\text{PL}})\right)$ using the Akaike information criterion \citep[AIC,][]{akaike_criterio} obtained from each PSD fit, and considering a p-value$\rm{\leq0.05}$.

We find the BPL fit to have a lower AIC than the PL fit with an RLM of $1.6\times 10^{-6}$, so the BPL better fits the PSD of our LC.

\begin{figure}[htp]
    \centering    \includegraphics[width=\columnwidth]{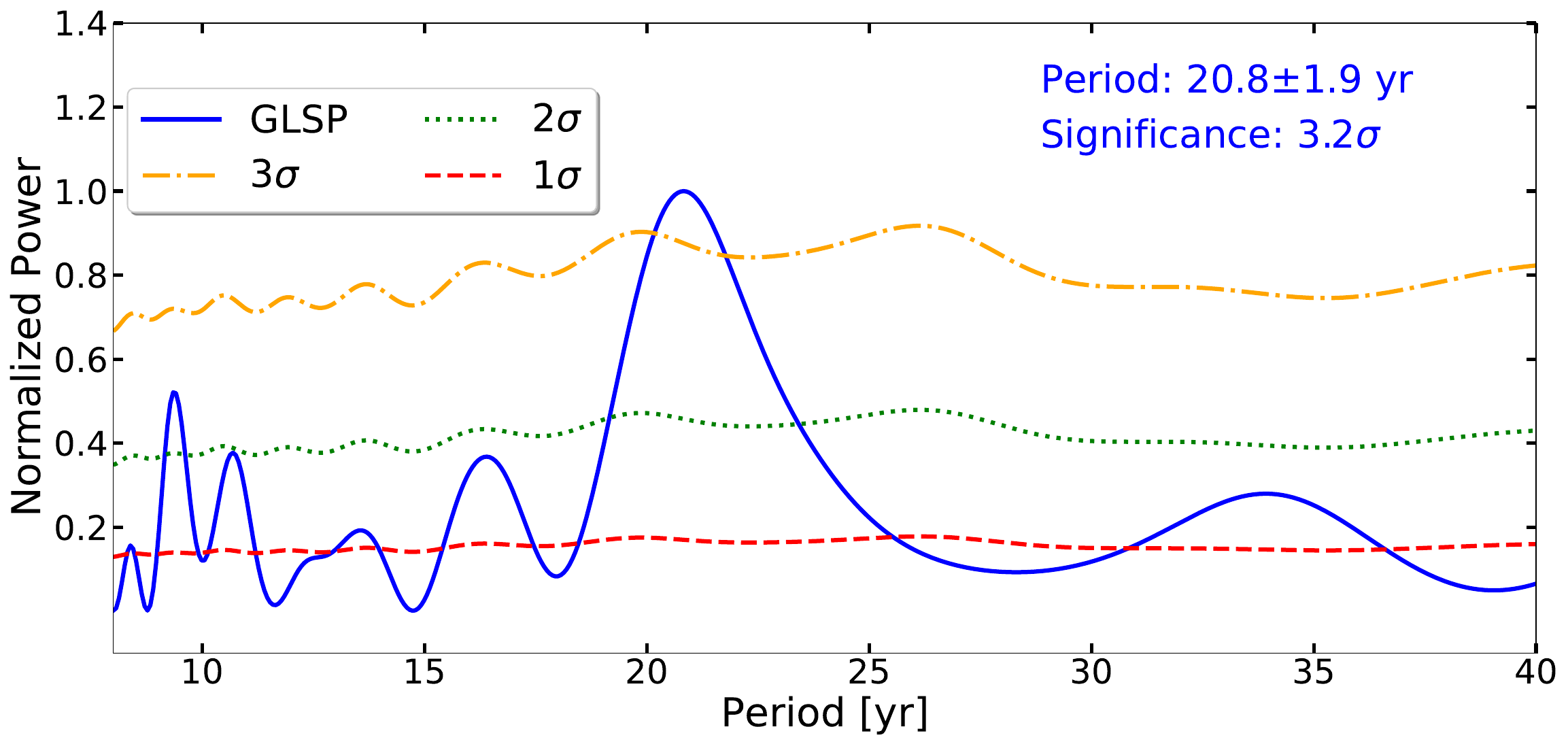}
    \caption{PG 1553+113 GLSP from the combined V-band LC. The highest peak is at 20.8$\pm$1.9 yrs. The significance is obtained with the PL model of $\S$\ref{sec: psd-estimation}.}
    \label{fig:pg53lsp}
\end{figure}

\begin{figure*}[htp]
    \centering
    \includegraphics[width=0.72\linewidth]{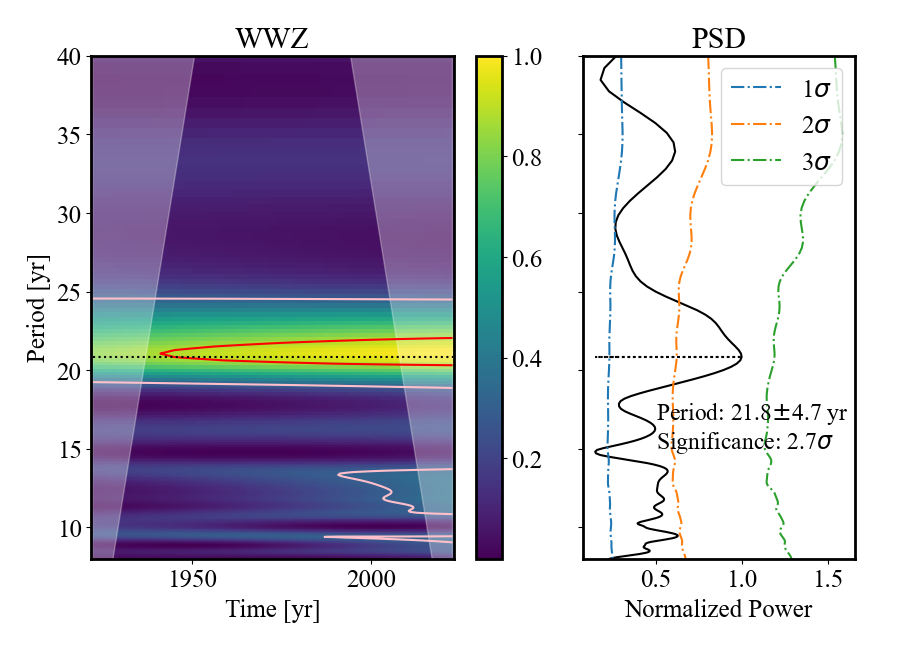}
    \caption{PG 1553+113 WWZ from the combined V-band LC. The color bar represents the normalized PSD, the pink and red contours represent $1\sigma$ and $2\sigma$ confidence, respectively, and the black dotted line shows the strongest period on both plots. A bright horizontal band covering the whole time scale suggests that the period of $21.8\pm4.7$ yrs is stable and present in the whole LC. The white-shaded region in the wavelet plot shows the region outside the cone of influence where the edge effects become important. The significance is obtained with the PL model of $\S$\ref{sec: psd-estimation}.}
    \label{fig:pg_wwz_lsp}
\end{figure*}

The outcome of the periodicity analysis is presented in Figures \ref{fig:pg53lsp} and \ref{fig:pg_wwz_lsp}. The GLSP reveals a longer-term period of $20.8\pm1.9$ yrs with a local significance (i.e.~not corrected for trials) of $3.2\sigma$ for the PL and $2.7\sigma$ for the BPL.

We use a Morlet wavelet with a decay constant $C=0.00058$, chosen to balance the number of data points in a wavelet with the WWZ implementation of \textit{libwwz} for our analysis. The WWZ analysis indicates a period of $21.8\pm4.7$ yrs, with a local significance of $2.7\sigma$ for the PL and $4.0\sigma$ for the BPL. This period is evident throughout the light curve (see Figure~\ref{fig:pg_wwz_lsp}). The $2\sigma$ contour level starts around 1950 and widens as we move forward in time which represents the change in data quality over time.
\subsection{Trial correction} \label{sec:trial_correction}
\begin{figure}
	\includegraphics[width=\columnwidth]{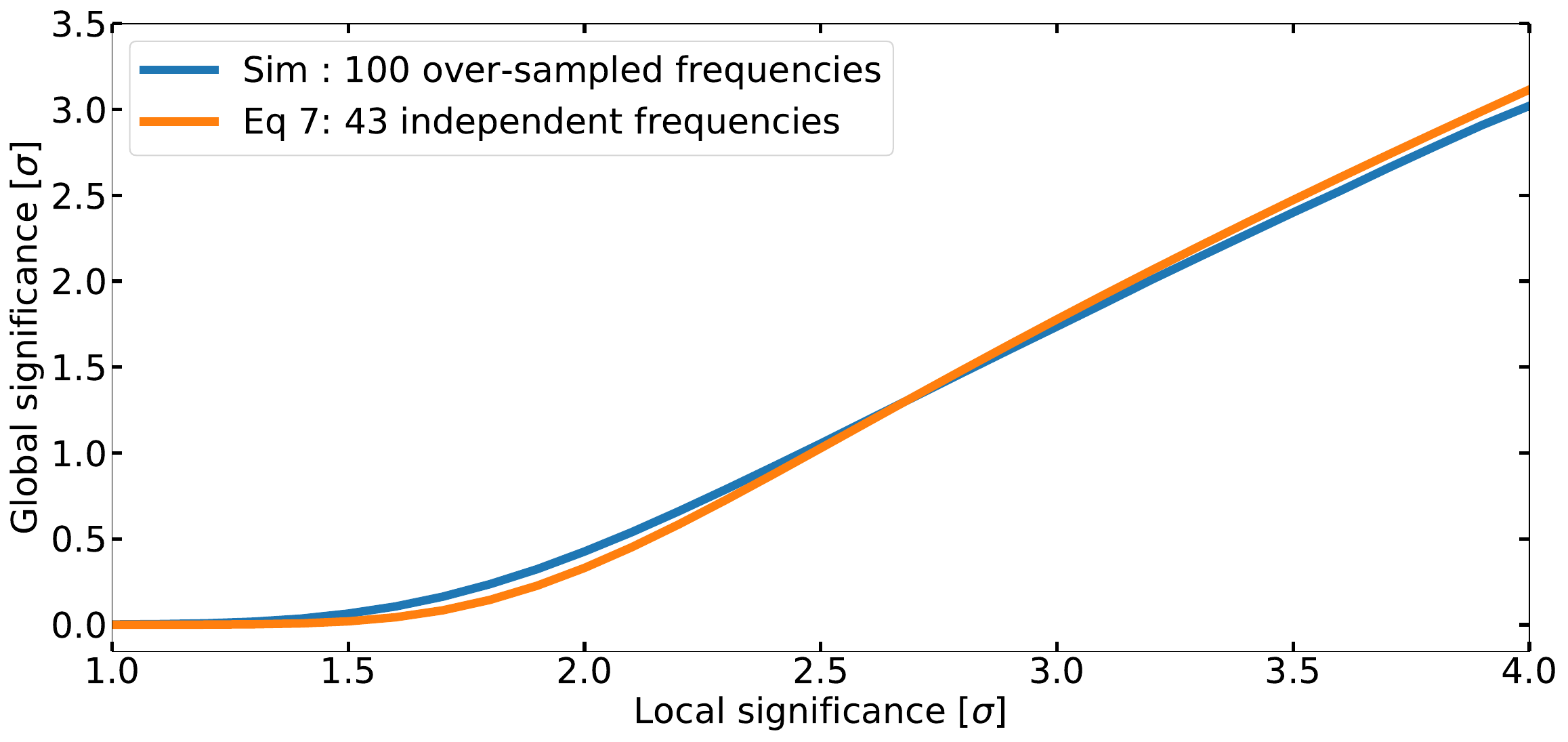}
	\caption{Relation to estimate the number of independent periods $N$ needed in equation~\ref{eq:trial} to compute the trial factor. ``Eq 7'' denotes the results of applying the equation \ref{eq:trial} for a specific number of independent frequencies.} \label{fig:trials}
\end{figure}

The significance obtained by the analysis methods has to be corrected for the look-elsewhere effect, resulting in the global significance of the period observed \citep{gross_vitells_trial}. This global significance is approximated by 
\begin{equation}\label{eq:trial}
p_{\mathrm{global}}=1-(1-p_{\mathrm{local}})^{N},
\end{equation}
where $p_{\mathrm{local}}$ is the local significance, and $N$ is the product of the number of sources searched and the number of independent periods in each periodogram \citep[in the current case, only one source was searched, c.f.][]{penil_2022}. $N$ is estimated by performing Monte Carlo simulations. Specifically, we apply the algorithm described in \citet{penil_2022}, using $10^{8}$ LCs simulated using the technique of \citet{timmer1995generating}. In this way, we obtain the experimental relationship between local and global significance considering an oversampling of 100 frequencies, and then we determine $N$ to be 43 by fitting Equation \ref{eq:trial} to the experimental relationship. We fit specifically over the range $2.5\sigma-3.0\sigma$, since that is the range of local significance we obtained in our analysis (see Figure \ref{fig:trials}).
Equation \ref{eq:trial} maps a local significance of $\approx3\sigma$ to a global significance of $\approx1.6\sigma$.

\subsection{False alarm probability}
As an alternative measure of significance, we determine the false alarm probability. This is the probability that a PG 1553+113-like, red-noise LC produces a peak period with a given local significance; we use 3$\sigma$ as a representative value of the local significances we found above in $\S$\ref{sec:results} \citep[e.g.,][]{scargle_1982, VanderPlas_2018}. We generate $10^6$ LCs using both the PL and BPL noise models with the fit parameters from $\S$\ref{sec: psd-estimation}, and use the same sampling as the 28-day binned LC of PG 1553+113. We find the false detection rate to be $\sim 4.9\%$ and $\sim 2.9\%$ for LCs generated using the PL and BPL noise models, respectively.
These values of false detection rate are similar to the $p$-value one would obtain from the $\sim1.6\sigma$ global significance we obtain in $\S$\ref{sec:trial_correction}. 

Based on the global significance, the false alarm rate for the 2.2-yr periodicity, as determined based on the findings of \citet{penil_mwl_pg1553}, is $\sim 0.25\%$ (corresponding to a global significance of $2.8\sigma$), indicating a low likelihood of spurious detection, and that for the 22-yr periodicity is $\sim 5.5\%$ (corresponding to a global significance of 1.6$\sigma)$. If the 2.2- and 22-yr periods are independent events, this implies that the false alarm probability for the simultaneous presence of both periods is roughly $\sim 0.014\%\ (3.6\sigma)$. As a corroboration of this estimate, we test the false alarm probability by simulating $10^6$ LCs and counting how many LCs show simultaneous periods of 2.2 and 22 yrs. For that purpose, we look for the 2.2-yr period in a range of $2.2\pm0.2$, and the 21.8-yr period in a range of $21.8\pm4.7$. The periods are searched within the same time range as the original analysis to maintain consistency: $2.2\pm0.2$ within the subrange of approximately 18 years, which aligns with MDB, and $21.8\pm4.7$ within the complete 100-year range. The resulting confidence level is $4.8\sigma$ (global significance of 3.8$\sigma)$\footnote{We estimated that the number of independent frequencies necessary to convert this local significance to a global one is 49.}, the highest significance with a million simulations for both PL and BPL noise models.

This low likelihood that both periods could arise stochastically motivates consideration of a model that predicts two periodicities; we explore such a model based on an accreting black hole binary in $\S$\ref{sec:discussion}, focusing on the modest question of whether two periods with a 10:1 ratio can arise in the gas dynamics. A complete model of the LC from such a system requires much more detailed predictions that are inaccessible with our simulations in $\S$\ref{sec:discussion}, such as the amplitude and shape of the periodic pulse trains, and the nature of a third, stochastic component (presumably with a similar physical origin as other blazars). We leave the development of such a complete model to future work, at which time we could make a meaningful assessment of its ability to account for all the variance in PG 1553+113's emission.
\subsection{Complementary study}
We perform further analysis to investigate how the DASCH, MDB, and DASCH+MDB segments of the LC have on our detection of the periodic signal. As previously discussed, the DASCH data has some limitations, including gaps in the LC and larger uncertainties in the data points compared to the MDB data.

Individual periodicity analyses are carried out for each subset of the LC separately, using the GLSP and WWZ methods (using the PL method to find the local significance). The results of this analysis are presented in Table~\ref{tab:partial_analysis}.
The GLSP and WWZ analyses of the DASCH data report significances 22\% and 15\% below that of the complete LC (DASCH+MDB), respectively. The MDB data does not show any meaningful period of around 22 yrs, which is expected since the temporal baseline of the MDB data is shorter than this period. These results indicate that the slow oscillation we report in this study is not solely due to the MDB but that the DASCH data plays a vital role in detecting this longer-term period of PG 1553+113.

\begin{table}[ht]
    \centering
    \caption{Periodicity analysis on LC subsets.}
    \label{tab:partial_analysis}
    \begin{tabular}{p{0.9cm}p{2.3cm}p{1.4cm}p{2.2cm}}
    \hline
     & DASCH & \centering MDB & {DASCH+MDB} \\
    \hline
    \hline
    GLSP & {$20.1\pm3.9$ yrs \,\, ($\approx2.5\sigma$)} &\centering\, {No detection} & {$20.8\pm 1.9$ yrs \,\, ($\approx3.2\sigma$)}\\
    \hline
    WWZ  & {$25.0\pm 12.5$ yrs ($\approx2.3\sigma$)} &\centering\, {No detection} & {$21.8\pm 4.7$ yrs \,\, ($\approx2.7\sigma$)}\\
    \hline
    \end{tabular}
\end{table}

Combining the new observations with the archived optical LC not only strengthens the power of the $\sim22$-yr peak, but also increases its significance by $\sim17\%-28\%$ and reduces the uncertainty by $\sim 60\%$.

We estimate how many more cycles of data we would need to achieve a $3\sigma$ global significance following the methods of \citet{penil_2022}. Assuming the quality of future data remains the same as the MDB data, we would need three more cycles (66 years) of observation to get a $3\sigma$ global significance against the PL noise model. For the BPL noise model, we require $\sim$ 40\% more observation than the PL noise model, which agrees with the fact that the PL noise model results in a higher estimate of the significance by $\sim$ 20\% compared to the BPL noise model.

\begin{figure}[htp]
    \centering
    \includegraphics[width=\columnwidth]{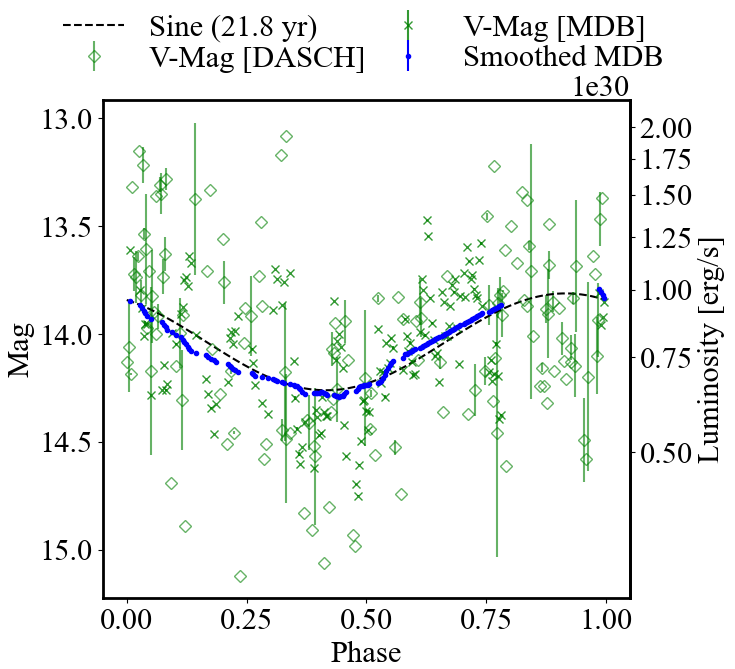}
    \caption{Phase folded plot with binned DASCH and MDB data. The smoothed MDB follows the 21.8 yr sinusoidal fit.}
    \label{fig:phase_compare}
\end{figure}

In Figure~\ref{fig:phase_compare}, we phase-folded the LC to highlight the oscillation, using the 21.8-yr period obtained from the WWZ method. We also smoothed the MDB data using the Savitzky-Golay filter \citep[][]{savitzky1964smoothing} with a second-order polynomial and a window length of 100, which reduces the scatter from the 2.2-yr periodicity and other stochastic components, revealing that the MDB LC closely follows the $\sim$22-yr oscillation.
\begin{figure*}[ht]
    \centering
    \includegraphics[width=\textwidth]{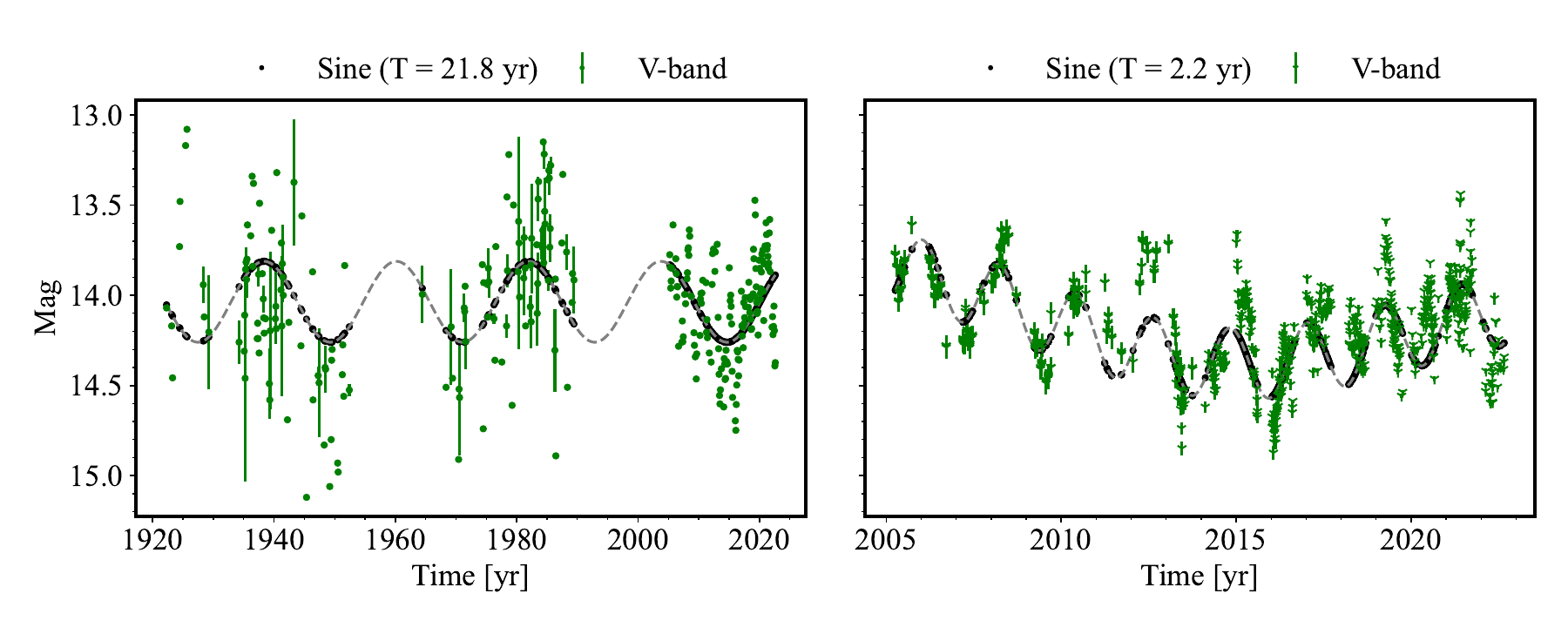}
    \caption{Sinusoidal reconstruction according to the two periods detected from the optical emission of PG 1553+113. \textit{Left}: a fit of the entire optical LC with a $\sim 22$-yr period. \textit{Right}: a fit of the MDB optical LC with a $\sim 2.2$-yr period. The $\sim 22$-yr fit is superposed, which explains the trends discussed in $\S$\ref{sec:intro}.}
    \label{fig:sin_fit}
\end{figure*}
\begin{figure}[h]
    \centering    \includegraphics[width=\columnwidth]{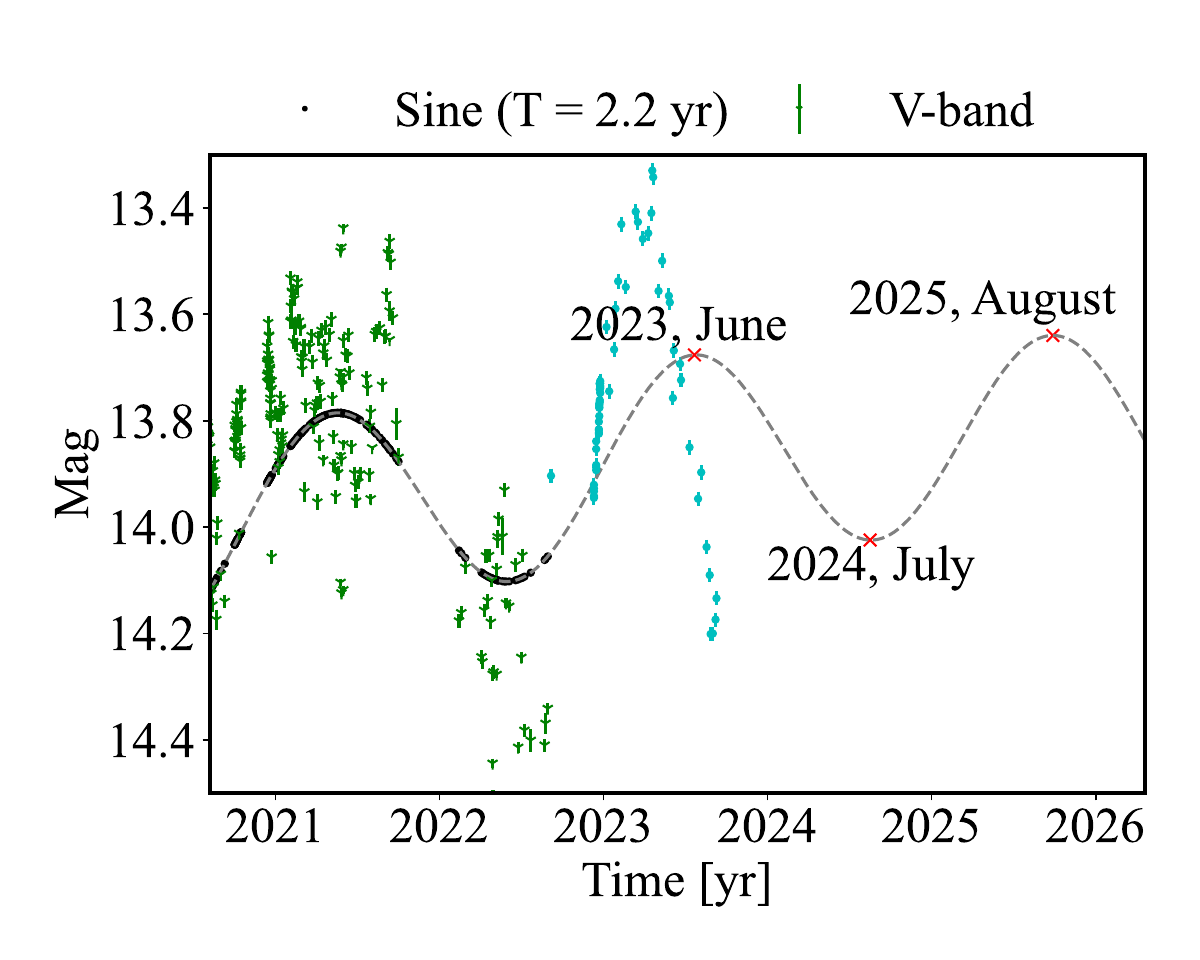}
    \caption{Sinusoidal fit from Figure~\ref{fig:sin_fit} (\textit{Right}) is extended further to predict the fast oscillation extrema. The next minima in optical emission is predicted around July 2024, and the next maxima in August 2025. More recent data points that are not analyzed in this work are shown in cyan. The maximum in recent data (19 April 2023) falls within the uncertainty of the fast oscillation estimation.}
    \label{fig:fast_lc_prediction}
\end{figure}
\begin{figure}[h]
    \centering    \includegraphics[width=\columnwidth]{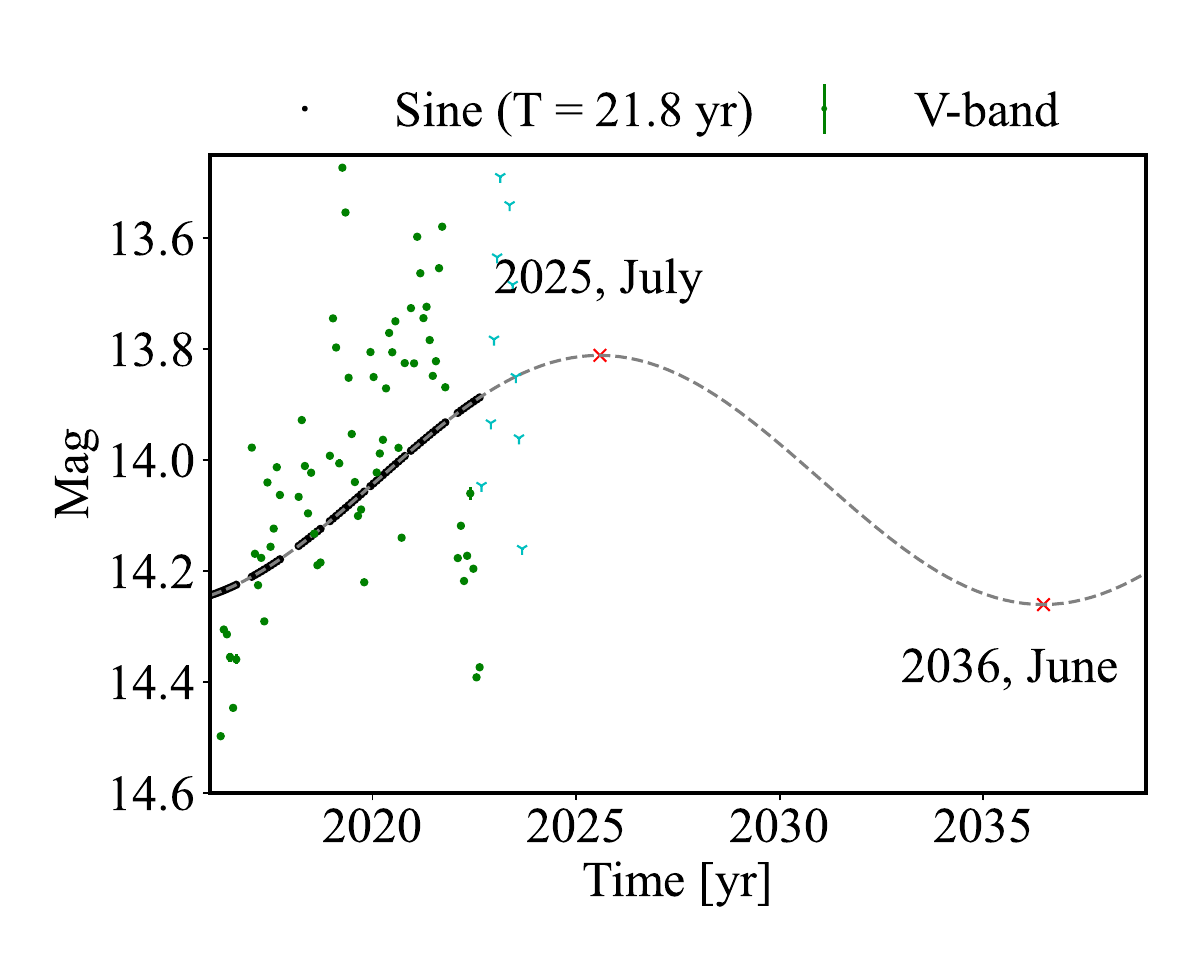}
    \caption{Sinusoidal fit from Figure~\ref{fig:sin_fit} (\textit{Left}) is extended further to predict the slow oscillation extrema. The next maximum in optical emission is predicted around July 2025, and the next minimum in June 2036.
    More recent data points that are not analyzed in this work are shown in cyan.}
    \label{fig:slow_lc_prediction}
\end{figure}

We present two sinusoidal reconstructions of the optical emission of PG 1553+113, one using the period of 21.8 yrs obtained from our analysis and the other using 2.2 yrs from the literature. These reconstructions are shown in Figure~\ref{fig:sin_fit}. We rely on the R-squared (R$^{2}$) criterion to assess the quality of both fits. R$^{2}$ is a statistical measure that assesses the goodness of fit of a regression model. Ranging from 0 to 1, higher values signify a better fit. Firstly, we estimate the R$^{2}$ for the individual sinusoid of 21.8 yrs over the entire LC (right panel of Figure~\ref{fig:sin_fit}), resulting in 0.5. Then, we estimate the R$^{2}$ for the reconstruction fit combining both oscillations of 21.8 yrs and 2.2 yrs over the MDB segment of the LC (left panel of Figure~\ref{fig:sin_fit}), resulting in 0.45. Both R$^{2}$ values are ``moderate" according to \citet{hair_r2_2011}. Additionally, the Ljung-Box test for the two fits reveals that the residuals are not independent. These findings suggest that the selected models may not fully capture certain aspects of the LC's structure. Specifically, this result could be interpreted as follows: while the sine model effectively captures the general periodic oscillations, it may overlook other factors, such as the shape of the oscillation or the changes in amplitude over time (as observed in the dispersion shown in Figure \ref{fig:fast_lc_prediction}) or the presence of noise.

\begin{figure}[ht]
    \centering
    \includegraphics[width=\linewidth]{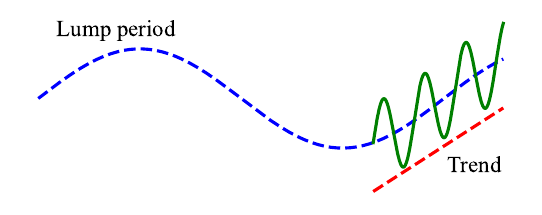}
    \caption{A trend in a PG 1553+113 LC, discussed in $\S$\ref{sec:intro}, is indicated in this illustration by the red dashed line. The fast and slow oscillations reflect the orbital and the lump periods, respectively, from our binary SMBH interpretation of the source (see $\S$\ref{sec:discussion}). Here, the trend is shown as a rising segment of the lump period.}
    \label{fig:trend_period}
\end{figure}

We forecast the future oscillations of both periodic oscillations. For the fast oscillation of 2.2 yrs, the next low-emission state is predicted to occur in July 2024, and the following high-emission state in August 2025 (see Figure~\ref{fig:fast_lc_prediction}). For this fast oscillation, no known delay between the optical and $\gamma$-ray emission exists \citep[both emissions are correlated, see][]{penil_mwl_pg1553}. Therefore, we expect the maxima/minima of $\gamma$-ray emission to co-occur with optical. We note a latest high emission in \textit{Fermi}-LAT on 21st April 2023, which is within the uncertainty of the prediction.\footnote{ \url{https://fermi.gsfc.nasa.gov/ssc/data/access/lat/LightCurveRepository/source.html?source_name=4FGL_J1555.7+1111}} For the slow oscillation of approximately 22 yrs (Figure~\ref{fig:slow_lc_prediction}), the next high-emission state is expected to occur in July 2025, followed by a low in June 2036.
In the absence of any lag, the $\gamma$-ray emission will occur concurrently with the optical counterpart.
We show a cartoon representing the double periodicity observed in the LC of PG 1553+113 (seen in Figure~\ref{fig:sin_fit}) in Figure~\ref{fig:trend_period}. The figure illustrates the long-term period plus a trend, discussed in $\S$\ref{sec:intro}, as part of a longer-term period that we have labeled the ``lump period" (see $\S$\ref{sec:discussion} for a discussion of the lump period).

The nature of the DASCH data prevents us from looking for the 2.2-yr period. This is due to the average number of observations in the archival data being 2.1 per year, compared to 8.2 per year in the MDB data. The limited number of observations and uneven sampling in the DASCH data make detecting such a short period challenging. The archival data, on average, just barely meets the theoretical sampling rate of 4 points per cycle \citep{landau1967sampling}, but to determine the frequency and amplitude of a signal, the sampling needs to be in phase with the signal. The observations in the DASCH data fail to meet this requirement.
\section{Discussion \& theoretical interpretation} \label{sec:discussion}
In this section, we assume a fiducial total black hole mass for PG 1553+113 (hereafter just ``PG 1553'') of $M=5\times10^8\, M_\odot$, where $M_\odot$ denotes a solar mass, and a binary semi-major axis $a=5\times10^{-3}\,$pc, which imply a binary orbital period $T_{\rm binary} = 1.5\,$yrs in the source frame. This orbital period corresponds to the observed $2.2$-yr periodicity when a redshift of $\sim 0.433$ is assumed \citep[][]{danforth+2010, johnson+2019, dorigojones+2022}. We denote the binary orbital frequency as $f_{\rm binary} \equiv 1/T_{\rm binary}$. A Newtonian treatment of the disk should suffice at distances $\simeq \mathcal{O}(a)$ from the black holes since the characteristic gravitational length scale $L=GM/c^2$ is much smaller than $a$, i.e.~$L/a \simeq 5\times10^{-3}$, where $G$ is Newton's gravitational constant and $c$ is the speed of light. This binary is likely in the gravitational wave-driven regime of orbital evolution \citep[][]{penil_mwl_pg1553}, and over the last 100 years, gravitational wave decay is on the order of $0.01\times T_{\rm binary}$ \citep[][]{peters1964}, which is negligible compared to our uncertainties. Our assumed mass is consistent with the estimate of $(0.4-8)\times10^8 M_\odot$, based on light curve variability \citep[][]{dhiman+2021}, and is typical of masses in existing binary models of PG 1553 \citep[][]{cavaliere+2017, tavani2018blazar, huang+2021}. See also \cite{ackermann2015multiwavelength} for a rough estimate of $1.6\times10^8 M_\odot$ based on the empirical relationship between blazar jet power and disk luminosity presented in~\cite{ghisellini+2014}. We adopt standard geometrically thin, optically thick, constant-$\alpha$ disk models \citep[][]{SS1973, goodman2003}, with a monatomic adiabatic equation of state (adiabatic index $\Gamma = 5/3$), a radiative efficiency $\eta=10\%$, and an opacity dominated by electron scattering. We use a physical radiative cooling prescription, and we relate the disk effective temperature $T_{\rm eff}$ to the midplane temperature $T$, surface density $\Sigma$, and opacity $\kappa$, via $T_{\rm eff}^4 = (4/3)\, T^4/(\kappa \Sigma)$ \citep[][]{FKR}. 

We perform two-dimensional viscous hydrodynamic simulations of binary accretion with an Eulerian grid-based code \texttt{Sailfish} \citep[for details, see][]{WS2022+}. Radiation pressure is omitted in simulations due to its technical difficulty. Instead, the characteristic aspect ratio $h$ of the disk is chosen to be similar to a disk whose radiation and gas pressures are accounted for. The aspect ratio is defined as $h=H/r$, where $H$ is the disk half-thickness. We consider a fiducial accretion rate $\dot{M} = 0.1 \dot{M}_{\rm Edd}$ typical of blazars \cite[][]{ghisellini+2014}, where $\dot{M}_{\rm Edd}\equiv L_{\rm Edd}/\eta c^2$ is the Eddington accretion rate and $L_{\rm Edd}$ is the Eddington luminosity. This fiducial accretion rate yields characteristic disk aspect ratios $h\in \lbrace 0.015, 0.03, 0.045 \rbrace$ when measured around an equilibrium circumsingle disk at radii $r\in \lbrace a, a/2, a/3 \rbrace$, respectively. Note that an accretion rate $\dot{M} = 0.1 \dot{M}_{\rm Edd}$ is consistent with PG 1553's putative binary being in the gravitational wave-driven regime \citep[][]{penil_2022}; below we describe the implications of this for the binary's eccentricity.

\subsection{Implications for the binary hypothesis} \label{sec:bhypo}
Accreting binaries are known to hollow out a cavity and generate an overdensity in the circumbinary disk \citep[][]{macfadyen+2008}, called a ``lump,'' for certain regions of parameter space \citep[][]{dorazio+2013, miranda+2017, duffell+2020, zrake+2021}, including disks initialized with a large misalignment relative to the orbital plane of the binary \citep[][]{moody+2019}. The lump period $T_{\rm lump}$ depends on the cavity size, and in~\cite{WS2022+} a range of $T_{\rm lump} = 4.5-10.8\times T_{\rm binary}$ is reported based on a large suite of radiatively cooled, adiabatic two-dimensional disk models around circular, equal-mass binaries. In that work, $T_{\rm lump}$ was found to depend most strongly on the disk's aspect ratio $h$; thinner disks exhibit larger $T_{\rm lump}$. This range for $T_{\rm lump}$ accommodates most values reported so far in the literature, with some exceptions extending as low as $T_{\rm lump}\sim3.6\times T_{\rm binary}$ \citep[][]{bowen_lump}. Hence, in this work, we consider our \emph{a priori} possible range to be $T_{\rm lump}=3.6-10.8\times T_{\rm binary}$.

Assuming PG 1553's $2.2$-yr periodicity corresponds to the cosmologically redshifted binary orbit $(1+z)T_{\rm binary}$, its $10\times$ longer period of $22$ yrs is consistent with the high end of this theoretical redshifted lump period range. The high end is mostly populated by two-dimensional models of adiabatic, radiatively cooled disks that are sufficiently thin \citep[$h \lesssim 0.09$, see, e.g.][]{farris_trend_binary, WS2022+}. Assuming a standard value for the dimensionless viscosity parameter of $\alpha=0.1$, PG 1553's disk aspect ratio satisfies this thinness condition on radial distances $r\sim \mathcal{O}(a)$ for accretion rates $\dot{M}\lesssim 0.6\times \dot{M}_{\rm Edd}$. BL Lac objects like PG 1553 are safely below such accretion rates \citep[see e.g.][]{paliya+2021}. Thus, the 10:1 period ratio exhibited in its optical light curve might be explained by a binary engine.

In binary accretion models that assume isothermal gas, lumps register in total black hole accretion rates when the binary has mass ratio $q\gtrsim 0.2$ and eccentricity $e\lesssim0.1$ \citep[][]{dorazio+2013, miranda+2017, munoz+2020, duffell+2020, zrake+2021}.
The same models predict that, in the gas-driven regime of orbital evolution, binaries accreting from prograde disks have an attractor value of eccentricity of a few tens of percent, with a wide basin of attraction \citep[e.g.][]{roedig_eccentricity1, roedig_eccentricity2, zrake+2021, dorazio+2021, siwek+2023}. One might therefore expect most binaries to have significant eccentricity upon entering the gravitational wave-driven regime, at which time the binary begins to circularize. On its face, since PG 1553's putative binary seems safely in the gravitational wave-driven regime, it is plausible that its eccentricity is now small enough ($e\lesssim0.1$) to generate a significant lump in its circumbinary disk.
If PG 1553 hosts a binary and a lump, then these constraints ($e\lesssim0.1$, $q\gtrsim 0.2$) can be applied to PG 1553's putative binary. In Section~\ref{sec:existingmodels}, we apply these constraints to some binary models proposed in the literature.

\subsection{How the lump can imprint on jet emission} \label{sec:howlump}
Since PG 1553 is a very high energy ($E>100\,$GeV) BL Lac blazar, its emission is overwhelmed by its jet(s). Thus, we must consider mechanisms by which lump periodicity could transmit to jet power. We consider two mechanisms, lump-modulation of the accretion rate onto the jet-launching black hole(s) (hereafter the ``accretion rate mechanism''), and lump-modulation of the supply of seed photons from the circumbinary disk (hereafter the ``seed photon mechanism''). The accretion rate in the disk is a systemic parameter in theoretical jet-launching mechanisms \citep[][]{blandford+1977, blandford+1982}, and observational evidence exists for a correlation between jet power and accretion \citep[e.g.][]{ghisellini+2014}. Therefore, one expects that accretion rate modulations imprint upon all components of jet spectral energy distributions (SEDs), so long as such modulations take place on time scales much longer than black hole dynamical times (so that jet power can adjust quasi-statically). This includes imprints on the synchrotron component, which dominates the optical band in two-component models of PG 1553's SED \citep[see, e.g.][]{osterman+2006, albert+2007, aleksic+2012, raiteri+2015, raiteri+2017}. 

On the other hand, thermal seed photons from the disk at a radius $\simeq 4.7\,a$ (presumed to be the characteristic lump location, yielding a lump orbital period of $\simeq 10 \times T_{\rm binary}$) peak in the range of mid-infrared to optical bands for a large range of accretion rates, i.e.~between $0.01-1\times \dot{M}_{\rm Edd}$. Such photons, upscattered by electrons with fiducial blazar jet Lorentz factors $\gamma \simeq 10$ \citep[][]{ghisellini+2014}, would reach at least soft X-ray energies. These are similar energetics of the external Compton process considered in~\cite{band+1986} (e.g.~optical seed photons upscattering to X-rays). Thus, since the $\sim22$-yr period is detected in the optical band, the lump model of this modulation indicates that the accretion rate mechanism must be operative. 

Even if a lump is present, how strongly lump periodicity transmits to black hole accretion rates can depend on model details, including physical parameters like the size of the black hole in relation to its minidisk \citep[e.g.~compare][]{farris_trend_binary, WS2022+}. Observations of lump periodicity at energies in the synchrotron bump could therefore shed light on details of the accretion physics.

While the presence of lump periodicity in optical emission implies that the accretion rate mechanism is operative, a comparison between optical and $\gamma$-ray bands shows tentative evidence that the seed photon mechanism is also operative. In the optical band, the most recent minimum of the $\sim 22$-yr period occurs around 2016. However, in the more limited temporal baseline of $\gamma$-ray observations from 2009 until now, although a rising trend is apparent, no obvious minimum occurs around 2016 \cite[][]{penil_mwl_pg1553}, although it is difficult to be sure without a longer temporal baseline. It appears as though lump periodicity in optical emission instead lags $\gamma$-rays by several years. Radio emission appears to share the same lag as $\gamma$-rays \cite[][]{penil_mwl_pg1553}. These lags might be accounted for in the lump model if the seed photon mechanism contributes significantly to $\gamma$-ray emission. Simulations predict that lump-modulated black hole accretion rates can lag lump-modulated thermal disk emission by $\approx20-30\%\times T_{\rm lump}$ \citep[see Figure~5 in][]{farris_trend_binary}. Thus, since $\gamma$-rays can receive imprints via both the accretion rate and seed photon mechanisms, whereas optical emission only receives an imprint via the accretion rate mechanism, it is possible that lump-induced minima in optical emission lag those in $\gamma$-ray emission. Continued monitoring of PG 1553 for many years is therefore crucially important for making an accurate determination of the relative phase of the $\sim22$-yr modulation between low- and high-energy bands.

It is important to note that the SED of PG 1553 is usually modeled using only synchrotron and synchrotron self-Compton components \citep[SSC models, see e.g.][]{osterman+2006, albert+2007, aleksic+2012, raiteri+2015, raiteri+2017}, and likewise for BL Lac objects generally \citep[but it has been cautioned that this may not always be justified, see e.g.][]{ghisellini+1996}. The adequacy of SSC fits of SEDs suggests that seed photons from the disk are negligible. However, since PG 1553 has a compton dominance of $\sim1$, a non-negligible contribution of external Compton emission may be present in its SED. Thus, a three-component SED model (SSC+EC) with a significant disk-driven external Compton component (which the multiwavelength data and binary hypothesis tentatively require) must appear degenerate with current SSC models to within uncertainties. Therefore, in future work it is well-motivated to carefully evaluate SSC+EC models of PG 1553.

\subsection{Implications for existing binary models} \label{sec:existingmodels}

\begin{figure*}[htp]
    \centering
    \includegraphics[width=\textwidth]{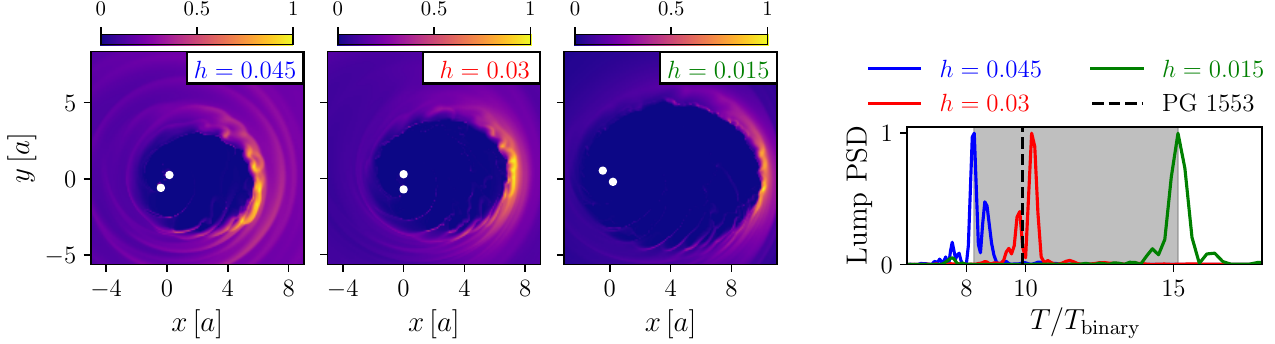}
    \caption{Simulations of accretion onto a circular binary with mass ratio $q=0.4$, total black hole mass $M=5\times10^8 M_\odot$, and orbital period of 1.5 yrs in the source frame. Three characteristic disk aspect ratios are shown, $h\in\{0.015, 0.03, 0.045\}$. The left panels show snapshots of the disk surface density on a linear color scale, normalized to their respective maxima. Black holes are indicated with white dots and are orbiting counter-clockwise. Thinner disks have larger cavities and correspondingly longer lump periods. A 10:1 relationship between lump and binary orbital periods is bracketed by these models (see the rightmost panel, where the shaded interval indicates the bracket spanning approximately 8:1 to 15:1). This is consistent with the 2.2-yr and 22-yr periods detected in PG 1553 light curves.}
    \label{fig:binary_models}
\end{figure*}

\cite{cavaliere+2017} and \cite{tavani2018blazar} propose a binary model of PG 1553's $2.2$-yr periodicity based on periodic driving of magnetohydrodynamic instabilities in the jet(s) due to the gravitational influence of the binary. The stipulated binary mass ratio is $q=0.1$, and a binary eccentricity of $e\sim0.2$ is obtained by fitting the model to PG 1553 light curves. Both of these parameters are outside the aforementioned lump-producing range suggested by isothermal binary accretion models. The model proposed in~\cite{cavaliere+2017} and~\cite{tavani2018blazar} is therefore disfavored under the hypothesis that the lump is causing the $22$-yr periodicity. It remains to be seen whether the binary parameters of this model can be adjusted into the lump-producing range while still producing an adequate fit of the $\sim 2.2$-yr periodicity.

Another recent binary model of PG 1553's $2.2$-yr periodicity, based on a two-jet precession scenario, arrives at $q\sim0.4$ and assumes $e=0$ \citep[][]{huang+2021}. These parameters fall into the aforementioned lump-producing range, and the mass ratio is consistent with one of the lower bounds ($q>0.35$) proposed by~\cite{caproni2017jet} based on jet precession scenarios. To assess whether this model can produce a 10:1 lump-to-binary period ratio, we ran $q=0.4$, $e=0$ binary accretion simulations with characteristic disk aspect ratios $h\in\lbrace 0.015, 0.03, 0.045 \rbrace$ (see our basic assumptions at the beginning of \sref{sec:discussion}). These disk aspect ratios are consistent with the lower bound $h > 0.0027-0.0048$ proposed in~\cite{caproni2017jet} based on jet precession models of the $\sim 2.2$-yr periodicity. We ran for $4000\times T_{\rm binary}$ at a resolution of $\Delta x = 0.016\,a$, corresponding to $1-10\times t_{\rm visc}$, where $t_{\rm visc} = (2/3) a^2/\bar{\nu}$ is a characteristic viscous time and $\bar{\nu}$ is the kinematic viscosity, then refined the grid to $\Delta x = 0.008\,a$ and ran for another $420\times T_{\rm binary}$. We perform analysis after the system has been allowed to settle for $100\times T_{\rm binary}$ post-grid refinement, i.e.~we perform analysis over the last $320\times T_{\rm binary}$. To quantify $T_{\rm lump}$, we compute the ``lump PSD,'' which we define as the PSD of the phase of the circumbinary disk's $m=1$ density moment -- i.e.~the PSD of the phase of $\int_0^{2\pi} \int_{1.5\,a}^{12\,a} \Sigma e^{im\theta} r dr d\theta$ with $m=1$; $T_{\rm lump}$ is then measured as the location of the lump PSD peak. This $m=1$ moment is computed over $1.5\,a < r < 12\,a$ in order to focus on the central black hole domain, and long-term trends due to cavity precession are subtracted before computing the PSD. Our simulation results are shown in Figure~\ref{fig:binary_models}. The first three panels show snapshots of $\Sigma$ on a linear color scale (normalized to their maximum values) for the three characteristic aspect ratios we consider, and the last panel shows the corresponding lump PSDs normalized to the lump peak. Differing cavity sizes are apparent in the snapshots, and the last panel shows that a 10:1 lump-to-binary period ratio is bracketed by these models (the bracket is indicated as the shaded interval, spanning between the lump PSD peaks from approximately 8:1 to 15:1). The binary parameters proposed in~\cite{huang+2021} therefore appear consistent with a lump model of PG 1553's $22$-yr periodicity. Furthermore, the theoretical uncertainty in the lump-to-binary period ratio, spanning at least from 8:1 to 15:1, could accommodate a considerable amount of statistical uncertainty in the measured 10:1 value. More complete thermodynamic modeling of the accretion disk (e.g.~including radiation and magnetic pressures), and bespoke constraints on PG 1553's accretion rate, are required to reduce the theoretical uncertainty. We are pursuing these improvements, and plan to report them in future work.

\subsection{Could the $22$-yr period be a minidisk evaporation cycle?}
Each black hole in an accreting binary has its own ``minidisk,'' while the whole binary is surrounded by a circumbinary disk. Recent numerical work, which includes radiation pressure from the inner disks around binary black holes, tentatively suggests that black hole minidisks can undergo cycles of formation and destruction on a time scale of $6-18\times T_{\rm binary}$ \citep[][]{williamson+2022}. In this process, as minidisks form, accretion onto the black holes ramps up, generating radiation pressure that eventually reaches a level that evaporates the minidisks. Once the minidisks evaporate, radiation pressure shuts off, allowing the minidisks to form again by feeding from the circumbinary disk.

A similar duty cycle of partial evaporation could conceivably cause accretion rate modulation on a time scale of $10\times T_{\rm binary}$. However,~\cite{williamson+2022} did not rule out the possibility that their proposed duty cycle is an artifact of manually and abruptly turning radiation pressure on and off. Furthermore, their setup assumed optically thin gas, appropriate for binaries with sufficiently wide separation. The gas at radii $r\sim a$ in PG 1553 should be highly ionized, and thus optically thick to electron scattering \citep[][]{haiman+2009}. Optical thickness would inhibit the ability of a central radiation source to evaporate the minidisks. Further theoretical work is therefore required to establish whether this process is realistic for a putative binary in a PG 1553-like system.

\subsection{Mechanism for PG 1553's $\sim2.2$-yr periodicity}

Stochastic flaring is an increasingly unlikely explanation of PG 1553's $2.2$-yr periodicity, since 7 or more putative periods are measured, depending on the energy band, and the statistical significance seems to be increasing over time \citep[see, e.g.][]{penil_2022, penil_mwl_pg1553}. The additional detection of a $22$-yr periodicity, consistent with a lump, further increases the chances that the $2.2$-yr periodicity is genuine. Two binary models of PG 1553's $2.2$-yr periodicity proposed in the literature are jet precession \citep[][]{caproni2017jet, huang+2021} and gravitationally induced magnetohydrodynamic instabilities in the jet(s) \cite[][]{cavaliere+2017, tavani2018blazar}. Since PG 1553's disk mass is expected to be negligible compared to the black hole mass, both of these mechanisms, both stemming from black hole masses and spins, are unlikely to be affected by the presence of a lump. However, accretion rate variability on the binary's orbital time scale might explain the $2.2$-yr periodicity, or will at least coexist with the aforementioned mechanisms. In this section, we enumerate accretion rate variability mechanisms that could explain PG 1553's $2.2$-yr periodicity, focusing on the time scales involved, and we judge their viability.

\begin{enumerate}
    \item Accretion variability at (or near) $f_{\rm binary}$: as is well-known \citep[e.g.~see the recent review in][]{lai+2022}, accretion rates are generally variable on the binary orbital time scale. In fact, such orbital peaks are usually found to be absent in only limited ranges of binary parameters, such as small mass ratios (i.e.~too small to form a lump), or towards the idealized circular, equal-mass case \cite[see, e.g.][]{macfadyen+2008, shi+2012, noble+2012, farris+2014, munoz+2020, duffell+2020}. 
    
    However, even for the circular, equal-mass case, it was recently found \citep[][]{WS2022+, WS+2023} that the minidisks can develop significant eccentricity, causing them to trade mass at a near-orbital beat frequency $f_{\rm binary} - f_{\rm prec}$ between the binary and the eccentric minidisk precession rate $f_{\rm prec}$. The generality of accretion variability at (or near) $f_{\rm binary}$, therefore, makes it a plausible candidate mechanism to generate the $2.2$-yr periodicity. 
    
    In addition, for mass ratios near $q\sim 0.4$, accretion rate variability at the lump and orbital periods were found to be comparable in recent work \citep[e.g.][]{duffell+2020, munoz+2020}. This suggests it is realistic to expect the observed similar amplitudes of the $2.2$-yr and $22$-yr periodicities in the PG 1553 optical light curve, but more realistic models are required to be sure.
    
    \item Accretion variability at $\sim (1.4-2)f_{\rm binary}$: two recent studies report mass transfer variability between minidisks, dubbed ``minidisk sloshing'' \citep[][]{bowen+2017, avara+2023}; the most recent study reports variability at a dominant frequency of $1.4f_{\rm binary}$. This effect was found to occur for sufficiently relativistic gravitational potentials (binary semi-major axis $a\lesssim 150 GM/c^2$), leading to associated black hole accretion rate variability. Since the putative binary in PG 1553 has $a \sim 200 GM/c^2$, this is unlikely to be the mechanism of the $2.2$-yr periodicity. If it were, then the $2.2$-yr period is less than a binary orbit, meaning the binary is even more widely separated (i.e.~less relativistic), further reducing the odds that the minidisk sloshing mechanism is operative.

    Accretion variability at $2f_{\rm binary}$ into the central cavity is also found for nearly circular, nearly equal-mass binaries \citep[e.g.][]{macfadyen+2008, farris+2014, miranda+2017}, likely due to the binary stripping material from the near-side of the cavity wall twice per orbit. However, such variability is generally found to be quite suppressed when measured at the black hole locations \citep[e.g.][]{farris+2014, munoz+2020, duffell+2020}, and therefore would not transmit significantly to a jet's base.
    
    \item Accretion variability at $2(f_{\rm binary} - f_{\rm lump})$, where $f_{\rm lump} \equiv 1/T_{\rm lump}$ \citep[e.g.][]{Roedig+2012, noble+2012, shi+2012, shi+2015}: for circular, equal-mass binaries with cavities that are sufficiently circular and close to the binary \cite[see exceptions in][]{farris_trend_binary, WS2022+, WS+2023}, variability was reported at a beat frequency between the binary and the lump, $2(f_{\rm binary} - f_{\rm lump})$. A 10:1 ratio between this frequency and $f_{\rm lump}$ is obtained if $T_{\rm lump} = 6\,T_{\rm binary}$, which is a theoretically reasonable value. While this is a possible mechanism to explain the $2.2$-yr variability, this operates in a narrow region of parameter space (i.e.~$e\sim0$, $q\sim1$, and small cavities), and thus is less general than e.g.~accretion variability at $f_{\rm binary}$.
\end{enumerate}

\subsection{Multiple periods from a single black hole} \label{sec:single}
It is known that electromagnetic emission from the vicinity of a single, accreting, spinning black hole can exhibit multiple QPOs simultaneously \citep[see][for a recent review]{ingram+2019}, although observing any QPOs at all from active galactic nuclei has been more difficult than from stellar-mass black holes \citep[][]{vaughan+2005}. These QPOs could arise from precessional phenomena driven by torques applied to the surrounding gas by the spinning black hole. For example, the relativistic precession model \citep[RPM, see, e.g.][]{stella+1998, stella+1999} predicts three hierarchical precessional frequencies for a particle in a geodesic orbit slightly off a compact object's equatorial plane. All three frequencies may be observed in rare cases \citep[e.g.][]{motta+2014}. More disk-like pictures involve the Bardeen-Petterson effect \citep[][]{bardeen+1975}, and for sufficiently large tilt angles between the disk and the black hole spin, the disk can ``break'' into two \citep[e.g.][]{lodato+2010} or more annuli that precess independently \citep[][]{liska+2021}. Such precession can steer a jet \citep[][]{liska+2018}, thereby imprinting periodic line-of-sight variation onto emission. In forthcoming work, we will present a detailed assessment of the single black hole precessional hypothesis for PG 1553's multiple periods.
\subsection{Characterization of gravitational wave signal and the merger time}
A binary SMBH system emits a gravitational wave (GW) signal until it coalesces \citep[][]{begelman1980massive}. We characterize the continuous GW frequency and strain amplitude emitted from PG 1553's putative binary central engine. The GW frequency is dominated by twice the orbital frequency, which corresponds to the continuous GW frequency of $\sim 4\times 10^{-8}$ Hz given the $1.5$-yr binary orbital period ($\sim 2.8\times 10^{-8}$ Hz in the observer frame, assuming a redshift $z=0.433$). Using the methods described by \citet{buskirk2019complete} and the implementation by \citet{2021JOSS....6.2968V}, we estimate the strain amplitude to be $\sim 10^{-17}$. In this estimate, we used a mass ratio of $q=0.4$, following \cite{huang+2021}. Our estimates of GW strain amplitude and frequency are similar to those reported in \citet{cavaliere2019supermassive}.

PG 1553's GW frequency is in the detectable range for pulsar timing arrays, but the strain is two orders of magnitude lower than the current sensitivity limit of NANOGrav \citep[see Figure~8c of][]{agazie2023nanograv}. Even with 20 years of observation, PG 1553's GW strain is expected to be $\sim10\times$ lower than the sensitivity of the Square Kilometer Array pulsar timing array \citep[][]{moore2014gravitational}.

The time to merger for a binary black hole system is given by $T_{\rm merge} = a^4 / 4\beta$ \citep[][]{peters1964}. Here $a$ is the semi-major axis and $\beta$ is defined as,
\begin{equation}
    \beta = \frac{64}{5} \frac{G^3}{c^5} m_1 m_2 (m_1+m_2),
\end{equation}
where $m_1=3.57\times10^8\, M_\odot$ and $m_2=1.43\times10^8\, M_\odot$ are the masses of two SMBHs. The merger time formula assumes a circular orbit, and gas-driving effects, while subdominant in the GW-dominated regime, are neglected completely in this merger time estimate. We find the $T_{\rm merge} \sim 15000$ yrs for the PG 1553+113 system. The estimation of merger time is sensitive to the total mass of the system; for the extremes of the total mass estimates $(0.4 - 8)\times10^8\, M_\odot$ \citep[][]{dhiman+2021}, the merger times range from $7000 - 10^6$ yrs. 

\subsection{Contrasting with OJ 287} \label{sec:oj287}
Motivated by existing models of PG 1553's 2.2-yr periodicity, in this work we followed suit by focusing on binary mass ratios near unity. In this section, we contrast this assumption with binary models of OJ 287, one of the most compelling binary blazar candidates to date.

Binary models of OJ 287 have a binary mass ratio of $q\sim 0.01$ \citep[see e.g.][]{valtonen+2012}, classified as an extreme mass ratio. A large eccentricity and inclination of the binary orbit with respect to the accretion disk is invoked, causing the secondary black hole to periodically punch through the disk twice in rapid succession, producing periodic double-peaked flares every $\sim12$ yrs. Misalignment of black hole spin with the orbital plane is also invoked to create relativistic precession of the binary orbit, in order to account for the presence of a second possible periodicity of 60 yrs.

Such a small mass ratio, large eccentricity, and inclined binary is not expected to produce a lump in the accretion disk of the primary black hole. Thus, the mechanism for two periodicities we considered in this work is not relevant in such models. Whether an OJ 287-like binary model could account for PG 1553's emission is a question we leave to future work.

\section{Summary} \label{sec:summary}
Motivated by the mounting evidence for a 2.2-yr QPO, suggestive of a binary engine, in this work we searched for additional QPOs in the historical record of PG 1553+113. Our analysis of the $\sim$100-yr V-band optical LC of PG 1553+113 from the DASCH database revealed hints of a multi-decade QPO of $\sim$22 yrs, which by itself has 1.6$\sigma$ significance after accounting for the look-elsewhere effect. Our estimate of the chance that colored noise produced both this oscillation and the known $\sim$2.2-yr oscillation, assuming they are independent occurrences, is ($3.6-3.8\sigma$). We performed an intense program of tests, including measuring the impact on the evidence for the $\sim$22-yr period from gaps in the data, the DASCH data alone, and the MDB data alone. We concluded it is unlikely that the $\sim22$-yr oscillation period is an artifact of such factors.

Recognizing the difficulty of producing a full model of a binary blazar LC, we instead explored the modest question of whether the gas dynamics around an accreting binary of SMBHs, with plausible PG 1553+113-like parameters, could exhibit two periodicites in a 10:1 ratio. Using two-dimensional viscous hydrodynamic simulations, we found that the so-called ``lump'' period in the circumbinary disk can exist in a wide range of ratios, spanning at least 8:1 to 15:1. This binary model, which requires a binary mass ratio $\gtrsim 0.2$ and eccentricity $\lesssim 0.1$ to generate a lump, can therefore accommodate the 10:1 ratio of QPO periods we report for PG 1553+113, and suggests a characteristic disk aspect ratio of $\sim$0.03. In this picture, the lump is an overdensity in the circumbinary disk that propagates along the cavity wall, modulating the rate of mass delivery to the binary on a $\sim$22-yr cycle. The jet(s) adjust quasi-statically to these modulations, leading to the observed, putative imprint of the $\sim$22-yr cycle. More detailed predictions of the jet emission from such a binary, such as the shape of each pulse train, their absolute and relative amplitudes, and the remaining component(s) of stochastic variability, are beyond the reach of the modeling we performed.


If the $\sim$22-yr QPO is genuine, we predict that its next peak will occur around July 2025. Our results might therefore contribute to our understanding of PG 1553+113 as a binary SMBH candidate. No confirmation of a binary from GWs can be expected, since the continuous GW signal from PG 1553+113's putative binary has a strain amplitude well below the detection threshold of foreseeable pulsar timing arrays \citep[][]{2021JOSS....6.2968V}.

To better understand periodicity in binary blazar emission, future theoretical work should constrain the relative importance of various mechanisms by which periodic processes can imprint on jet emission, such as the aforementioned accretion rate and seed photon mechanisms \citep[e.g.,][]{penil_mwl_pg1553}, jet precession \citep[e.g.,][]{huang+2021}, and gravitationally induced magnetohydrodynamic instabilities \citep[e.g.,][]{cavaliere+2017}. A greater understanding of such processes may teach us more about the binary scenario, and jet physics generally, using information we did not evaluate in detail, such as the relative amplitude of the $2.2$-yr and $22$-yr QPOs, and multiwavelength data.
 
\software{
	astropy \citep{astropy_2013, astropy_2018, astropy:2022}, 
	emcee \citep {emcee}, 
	Simulating light curves \citep{connolly_code},
    Simulating gravitational waves \citep{2021JOSS....6.2968V},
    WWZ \url{https://github.com/RedVoxInc/libwwz},
	Riroriro \url{https://github.com/wvanzeist/riroriro}
}
\section{Acknowledgments}
We want to thank all the observatories from which we used data. The DASCH project at Harvard is grateful for partial support from NSF grants AST-0407380, AST-0909073, and AST-1313370. We thank the Las Cumbres Observatory and its staff for their continuing support of the ASAS-SN project. ASAS-SN is supported by the Gordon and Betty Moore Foundation through grant GBMF5490 to the Ohio State University and NSF grants AST-1515927 and AST-1908570. Development of ASAS-SN has been supported by NSF grant AST-0908816, the Mt. Cuba Astronomical Foundation, the Center for Cosmology and Astroparticle Physics at the Ohio State University, the Chinese Academy of Sciences South America Center for Astronomy (CAS-SACA), the Villum Foundation, and George Skestos. The AAVSO database: Kafka, S., 2021, Observations from the AAVSO International Database, \url{https://www.aavso.org}. The CSS survey is funded by the National Aeronautics and Space Administration under Grant No. NNG05GF22G issued through the Science Mission Directorate Near-Earth Objects Observations Program. The Catalina Real-Time Transient Survey is supported by the U.S.~National Science Foundation under grants AST-0909182 and AST-1313422. Based on observations obtained with the Samuel Oschin Telescope 48-inch and the 60-inch Telescope at the Palomar Observatory as part of the Zwicky Transient Facility project. ZTF is supported by the National Science Foundation under Grant
No. AST-2034437 and a collaboration including Caltech, IPAC, the Weizmann Institute for Science, the Oskar Klein Center at
Stockholm University, the University of Maryland, Deutsches Elektronen-Synchrotron and Humboldt University, the TANGO
Consortium of Taiwan, the University of Wisconsin at Milwaukee, Trinity College Dublin, Lawrence Livermore National
Laboratories, and IN2P3, France. Operations are conducted by COO, IPAC, and UW.

P.P. and M.A. acknowledge funding under NASA contracts 80NSSC20K1562 and 80NSSC24K0635. 

S.B. acknowledges financial support by the European Research Council for the ERC Starting grant MessMapp, under contract no. 949555.

A.D. is thankful for the support of the Ram{\'o}n y Cajal program from the Spanish MINECO, Proyecto PID2021-126536OA-I00 funded by MCIN / AEI / 10.13039/501100011033, and Proyecto PR44/21‐29915 funded by the Santander Bank and Universidad Complutense de Madrid.

A.R. acknowledges the support of an Investigo grant funded by the European Union, Next Generation EU.
\bibliography{literature.bib} 
\bibliographystyle{apj}
\end{document}